\documentclass[12pt,a4paper]{article}
\usepackage{amsmath,amsfonts,amssymb,amsthm}
\usepackage{graphicx}
\usepackage{mathpple}
\newcommand\bp{{\mathbf p}}

\newcommand\bI{{\mathbf I}} 
 
\newcommand\bK{{\mathbf K}} 
\newcommand\bL{{\mathbf L}} 
\newcommand\bM{{\mathbf M}} 
\newcommand\bN{{\mathbf N}}
  
\newcommand\bP{{\mathbf P}}
\newcommand\bQ{{\mathbf Q}} 
 
\newcommand\bS{{\mathbf S}}

\newcommand\efs{{\mathfrak s}}

\newcommand\cA{{\mathcal A}}

\newcommand\cG{{\mathcal G}}

\newcommand\cM{{\mathcal M}} 
\newcommand\cN{{\mathcal N}}
  
\newcommand\cP{{\mathcal P}}

\newcommand\cT{{\mathcal T}}

\newcommand\mvector{\boldsymbol}

\newcommand\va{\mvector{a}}

\newcommand\ve{\mvector{e}}

\newcommand\vl{\mvector{l}}

\newcommand\vn{\mvector{n}}

\newcommand\vp{\mvector{p}}
\newcommand\vq{\mvector{q}}

\newcommand\vs{\mvector{s}}
\newcommand\vt{\mvector{t}}

\newcommand\field{\mathbb}
\newcommand\R{\field{R}}

\newcommand\C{\field{C}}
\newcommand\Z{\field{Z}}
\newcommand\N{\field{N}}

\newcommand\bbP{\mathbb{P}}
\newcommand\CP{\C\bbP^1}

\newcommand\bOmega{\boldsymbol{\Omega}}

\newcommand\diag{\operatorname{diag}}

\renewcommand\Re{\operatorname{Re}}

\newcommand\ord{\operatorname{ord}}

\newcommand\rmd{\mathrm{d}}

\newcommand\rmi{\mathrm{i}}
\newcommand\rme{\mathrm{e}}
\newcommand\Dt{\frac{\rmd\phantom{t} }{\rmd t}}

\newcommand\cn{\operatorname{cn}}
\newcommand\sn{\operatorname{sn}}
\newcommand\dn{\operatorname{dn}}

\usepackage[square]{natbib}
\theoremstyle{plain}
\newtheorem{theorem}{Theorem}
\newtheorem{lemma}{Lemma}
\theoremstyle{definition}
\newtheorem{example}{Example}
\newtheorem{remark}{Remark}
\author{Andrzej J.~Maciejewski,\\
  Institute of Astronomy,
  University of Zielona G\'ora \\
  Podg\'orna 50, PL-65--246 Zielona G\'ora, Poland,\\
  (e-mail: maciejka@astro.ia.uz.zgora.pl)       \and
  Maria Przybylska \\ 
  INRIA Projet \textsc{Caf\'e} \\
  2004, Routes des Lucioles, B.~P. 93 \\
  06902 Sophia Antipolis Cedex, France,  \\
  and \\
Toru\'n Centre for Astronomy,
  Nicholaus Copernicus University, \\
  Gagarina 11, PL-87--100 Toru\'n, Poland \\
  (e-mail: Maria.Przybylska@sophia.inria.fr) 
} 
\title{\hfill{\large\textsf{to appear in Celestial Mechanics}}\\[3cm] Non-integrability of the problem of a rigid satellite in
  gravitational and magnetic fields}
\begin{document}

\maketitle
\begin{abstract}
  In this paper we analyse the integrability of a dynamical system
  describing the rotational motion of a rigid satellite under the
  influence of gravitational and magnetic fields. In our
  investigations we apply an extension of the Ziglin theory developed
  by Morales-Ruiz and Ramis. We prove that for a symmetric satellite
  the system does not admit an additional real meromorphic first
  integral except for one case when the value of the induced magnetic
  moment along the symmetry axis is related to the principal moments
  of inertia in a special way.
\end{abstract}

\section{Introduction}
\label{sec:intro}

\begin{figure}[th]
\label{fig:model}
\centering \includegraphics[scale=0.8]{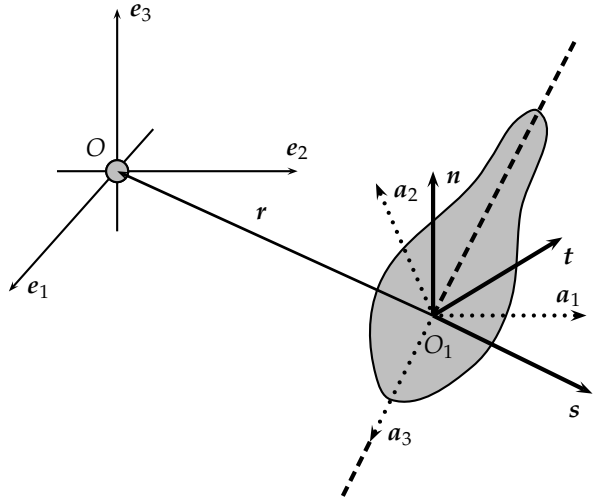}
\caption{A rigid satellite in an orbit around a gravitational centre}
\label{fig:sat}
\end{figure}

Let us consider a rigid body $\mathcal{B}$ with  mass $m$ and 
centre of mass $O_1$ moving in the gravitational field of a point $O$
with  mass $M$, see Fig.\ref{fig:sat}. We assume that the orbit is
circular and that it lies in the $(x,y)$-plane in the inertial reference
frame defined by the orthonormal versors $\{\ve_1,\ve_2,\ve_3\}$ with the
origin at $O$. The principal axes reference frame of the body with the
origin at $O_1$ is given by the orthonormal versors
$\{\va_1,\va_2,\va_3\}$. We describe the rotational motion of the body
with respect to the orbital reference frame $\{\vs,\vt, \vn\}$ with
the origin at $O_1$. Its axes lie along the radius vector of the
centre of mass of the body, the tangent to the orbit in the orbital
plane, and the normal to the orbital plane, respectively.
%
%

We accept the following convention, see \cite{Arnold:78::}. For a
vector $\vq$ we denote by $\bQ=[ Q_1, Q_2, Q_3]^T$  the associate
coordinates in the body frame, i.e., $Q_i = \va_i\cdot\vq$, for $ i =
1,2,3$. For two vectors $ \vq$ and $\vp$ we denote  their scalar and vector
products by $\vq\cdot\vp$ and $\vq\times\vp$, expressed in terms of their
coordinates in the body frame by $\langle \bQ,\bP\rangle$,
and $[\bQ,\bP]$, respectively. Thus we have
\begin{equation*}
 \langle\bQ,\bP\rangle := \sum_{i=1}^3Q_iP_i=\bQ^T\bP=\vq\cdot\vp, 
\end{equation*}
and
\begin{equation*}
[\bQ,\bP]:=\begin{bmatrix}
 Q_2P_3 -Q_3P_2\\
 Q_3P_1 -Q_1P_3\\
 Q_1P_2 -Q_2P_1
\end{bmatrix}=
\begin{bmatrix}
(\vq\times\vp)\cdot\va_1\\
(\vq\times\vp)\cdot\va_2\\
(\vq\times\vp)\cdot\va_3
\end{bmatrix}.
\end{equation*}

The equations of the rotational motion of the body can be
written in the following form
\begin{equation}
 \label{eq:euler}
 \begin{split}
 \Dt\bM &=[\bM,\bOmega]+ \bP,\\
 \Dt\bN&=[\bN, \bOmega],\\
 \Dt \bS&=[\bS,\bOmega-\omega_{\mathrm{O}} \bN],
 \end{split}
\end{equation}
where $\bM$, $\bOmega:= \bI^{-1}\bM$, $\bI:=\diag(A,B,C)$ are
the angular momentum, the angular velocity and the inertia tensor of
the body, respectively; $\omega_{\mathrm{O}}$ denotes the orbital
angular velocity of the  centre of mass of the body and $\bP$ is the
torque acting on the body.  The explicit form of $\bP$ depends on
a particular model. The gravity-gradient torque is usually approximated by 
the following formula
\begin{equation*}
\bP_{\mathrm{G}}:= 3\omega_{\mathrm{K}}^2 [\bS,\bI\bS],
\end{equation*}
where
\[
\omega_{\mathrm{K}}^2 = \frac{GM}{r^3},
\]
and $r$ is the radius of the orbit, see
\cite{Beletskii:65::,Beletskii:75::,Duboshin:68::}. Let us note that
in the case of a circular Keplerian orbit $\omega_{\mathrm{O}}=
\omega_{\mathrm{K}}$. Examples of models with $\omega_{\mathrm{O}}\neq
\omega_{\mathrm{K}}$ can be found in
\cite{Maciejewski:97::c,Maciejewski:01::j}.

In this paper we consider the case when, in addition to the
gravitational torque, also the magnetic torque plays a significant role.
Namely, we assume that the gravity centre (the Earth) is the source of
a magnetic field which can be well approximated by a magnetic dipole
whose axis coincides with $\ve_3$. Modelling of the magnetic torque
$\bP_{\mathrm{M}}$ is generally difficult because it depends not only
on the presence of constant magnets located in the satellite, but 
also  on magnetic and conductive properties of the material
used for its construction, as well as on the presence of electronic
equipment, for details see \cite{Beletskii:85::}. In this paper we
assume that the magnetic moment of the satellite is induced by the
magnetic field of the central body, and, moreover,   that the
body is magnetically symmetric along an axis $\vl$ fixed in the body.
Then we have
\begin{equation*}
\bP_{\mathrm{M}}:= \xi \langle \bL,\bN \rangle [\bL,\bN],
\end{equation*}
where $\xi$ is a parameter depending on the strength of the central
magnetic dipole and magnetic properties of the body.

Thus, we consider the following system  
\begin{equation}
 \label{eq:bel}
\left.
 \begin{split}
 \Dt\bM &=[\bM,\bOmega]+  3\omega_{\mathrm{K}}^2 [\bS,\bI\bS] + 
\xi \langle \bL,\bN \rangle [\bL,\bN],\\
 \Dt\bN&=[\bN, \bOmega],\\
 \Dt \bS&=[\bS,\bOmega-\omega_{\mathrm{O}} \bN].
 \end{split}\qquad \right\}
\end{equation}
It  possesses the Jacobi type first  integral
\begin{equation}
\label{eq:ener}
H=\frac{1}{2}\langle\bM,\bI^{-1}\bM\rangle-
 \omega_{\mathrm{O}}\langle\bM,\bN\rangle+
\frac{3}{2} \omega_{\mathrm{K}}^2\langle\bS,\bI\bS\rangle
-\frac{1}{2}\xi \langle \bL,\bN \rangle^2,
\end{equation}
and three geometric first integrals
\begin{equation}
\label{eq:geo}
H_2=\langle\bS,\bS\rangle,\qquad
H_3=\langle\bN,\bN\rangle,\qquad
H_4=\langle\bN,\bS\rangle.
\end{equation}  
The above equations can be rewritten in the Hamiltonian form
\begin{equation}
\label{eq:hform}
\Dt M_i=\{M_i,H\},\qquad 
\Dt N_i=\{ N_i,H \},\qquad
\Dt S_i=\{S_i,H\},\qquad i=1,2,3,
\end{equation} 
where the Poisson bracket $\{\cdot,\cdot\}$ is defined by
\begin{equation}
\label{eq:poiss}
\begin{split}
\{M_i,M_j\}&=-\sum_{k=1}^3\varepsilon_{ijk}M_k,\qquad
\{M_i,N_j\}=-\sum_{k=1}^3\varepsilon_{ijk}N_k,\\
\{M_i,S_j\}&=-\sum_{k=1}^3\varepsilon_{ijk}S_k,\qquad
\{N_i,N_j\}=\{S_i,S_j\}=\{N_i,S_j\}=0,
\end{split}
\end{equation}
where $\varepsilon_{ijk}$ is the Levi-Civita symbol. This
Poisson bracket is degenerated and the three geometric integrals
\eqref{eq:geo} are its Casimirs.  Their common levels are symplectic
manifolds \cite{Marsden:99::}.  From the geometric interpretation of the 
vectors $\bN$ and $\bS$ it follows that, for further study, we can select
the following six dimensional symplectic leaf
\begin{equation}
\label{eq:sympl}
 \cM^6=\{(\bM,\bN,\bS)\in\R^9 \,|\,\,  \langle\bS,\bS\rangle=1,\quad
\langle\bN,\bN\rangle=1,\quad
\langle\bN,\bS\rangle=0  \}, 
\end{equation}
which is diffeomorphic to $\R^3\times \mathrm{SO}(3,\R)$. 
\begin{remark}
\textsl{
  The configuration space of a rigid body whose centre of mass moves
  in a prescribed orbit is $\mathrm{SO}(3,\R)$ --- all possible
  orientations of the body with respect to the orbital frame. Thus the
  classical phase space of the system is $T^*\mathrm{SO}(3,\R)\simeq
  \R^3\times\mathrm{SO}(3,\R)$.}
\end{remark} 
\begin{remark}
\textsl{
  We can look at system~\eqref{eq:bel} as a Hamiltonian system defined
  on a nine dimensional Poisson manifold which is $\efs^*$ --- the dual
  to  nine dimensional Lie algebra $\efs = (\R^3\times\R^3)\rtimes
  \mathrm{so}(3,\R)$ (here $\rtimes$ denotes the semi-direct product
  of Lie algebras). Then the Poisson bracket defined
  by~\eqref{eq:poiss} is the standard Berezin-Kostant-Kirillov-Souriou
  bracket, and $\cM^6$ is a co-adjoint orbit, see \cite{Marsden:99::}.
  Here we refer the reader to paper~\cite{Audin:02::} where the
  case of a rigid satellite without the influence of magnetic torques is
  considered.}
\end{remark} 
System \eqref{eq:bel} depends on the parameters $p:= (A,B,C,
\omega_{\mathrm{O}}, \omega_{\mathrm{K}}, L_1,L_2,L_3,\xi)$. They
belong to  a set
\begin{equation*}
\cP:=\{p \in\R_+^5\times\R^4 \;|\; \langle \bL, \bL \rangle = 1, \quad
A< B+C, \quad B<C+A, \quad C<A+B, \},
\end{equation*}
whose interior is an eight-dimensional subset of $\R_+^5\times\R^4$
($\R_+$ denotes the positive real axis).

It is natural to ask for which $p\in\cP$ system \eqref{eq:bel} or its
restriction to $\cM^6$ admits one or two additional first integrals.
The high dimensionality of the system and a big number of parameters
make this problem very difficult. Let us enumerate some known  facts.
\begin{enumerate}
\item For $\xi=0$ (the magnetic torque vanishes) the only known
  completely integrable case is  a spherically symmetric case
  $A=B=C$.  This case is trivial because for a spherically symmetric
  body the gravitational torque vanishes. There is no proof that
  system \eqref{eq:bel} is non-integrable when $\xi =0$ and the body
  is not spherically symmetric.
\item For $\xi=0$ and an axially symmetric body, e.g. $A=B$,
  system~\eqref{eq:bel} admits one additional first integral, namely
  $H_5 = M_3$. There is no proof that this is the only situation when
  system~\eqref{eq:bel} possesses one additional first integral.
\item For $A=B=C$ only the magnetic torque acts on the body.
  System~\eqref{eq:bel} is completely integrable and the additional first
  integrals are $H_5=\langle \bM, \bN\rangle$ and $H_6=\langle \bM,
  \bL\rangle$. In this case the first two equations form a closed
  subsystem which coincides with a special case of the Kirchhoff
  equations for a rigid body in ideal fluid in the integrable case of
  Clebsh, see \cite{Kozlov:96::}.
\end{enumerate}
Some limiting cases of system~\eqref{eq:bel} when
$\omega_{\mathrm{O}}=0$, or $ \omega_{\mathrm{K}}=0$ are worth
mentioning because they are related to  very well known systems.

Let us consider the case $\omega_{\mathrm{O}}=0$. Now,
system~\eqref{eq:bel} describes the rotational motion of a rigid body
with the mass centre fixed in the external gravity and magnetic
fields. For $\xi=0$ the first and the third equation in~\eqref{eq:bel}
form a closed subsystem which coincides with the equations of motion
of the completely integrable Brun problem \cite{Brun:1893::}, see also
\cite{Bogoyavlenskii:85::a}.  When $\omega_{\mathrm{K}}= 0$, a
subsystem of~\eqref{eq:bel} consisting of the first two equations, is again
a special case of the Kirchhoff equations, see \cite{Kozlov:96::}.

The aim of this paper is to study the integrability of
system~\eqref{eq:bel} when the body is axially symmetric. For this
purpose we apply the Morales-Ramis theory
\cite{Morales:99::,Morales:01::b} which is an extension of the Ziglin
theory \cite{Ziglin:82::b,Ziglin:83::b}.  Both theories are based on a
study of variational equations around a particular non-equilibrium
solution of the complexified system. We can associate with the
variational equations the monodromy and the differential Galois
groups. When the system is integrable, then these groups are of a
special form and this fact gives a necessary condition for
integrability.  To make the paper self-contained, we present basic
theoretical facts concerning the Ziglin and Morales-Ramis theory in
the next section.  More technical material needed in our investigation
is presented in the Appendix. We present both theories trying to avoid
formal language, and we give several examples, which, as we hope,
helps to understand basic notions of both theories and to popularise
them in the celestial mechanics community. It is worth mentioning that
one of the most difficult problems of celestial mechanics---the
question about the non-integrability of the three-body problem---has
been recently solved with the help of these theories, see
\cite{Tsygvintsev:00::a,Tsygvintsev:01::b,Tsygvintsev:01::a,Boucher:00::}.
We remark here that H.~Poincar\'e~\cite{Poincare:1890::} investigated
the question of integrability of the three-body problem however he
assumed that the first integrals are holomorphic functions of the
perturbation parameter (mass of one body). Thus, his non-integrability
theorems do not assert anything for fixed value of this parameter. 

In Section~\ref{sec:var} we derive the variational equations along a
family of particular solutions. Our first non-integrability theorem is
formulated and proved in Section~\ref{sec:non}. We show in this
section that the complexified system considered does not possess an
additional \emph{complex meromorphic} first integral which is
functionally independent from the Hamiltonian. The question whether
the system does not possess an additional \emph{real meromorphic}
 first integral is much more difficult. We investigate it in
the last Section combining the differential Galois approach with
the Ziglin argumentation~\cite{Ziglin:97::}.

\section{Theory}
\label{sec:th}
In this section we describe informally basic facts concerning the
Ziglin and Morales-Ramis theories. For detailed exposition we refer
the reader to \cite{Braider:96::,Morales:99::,Audin:01::}.

Let us consider a complex dynamical system 
\begin{equation}
\label{eq:ds}
\Dt x = v(x), \qquad t\in\C, \quad x\in M^n, 
\end{equation}
where $M^n$ is a complex $n$-dimensional analytic manifold (we can think
that $M^n$ is just $\C^n$).  If $\varphi(t)$ is a non-equilibrium
solution of \eqref{eq:ds}, then the maximal analytic continuation of
$\varphi(t)$ defines a Riemann surface $\Gamma$ with $t$ as a local
coordinate.  Here it is important to distinguish between the abstract
Riemann surface $\Gamma$ and its image $i(\Gamma)$ in $M^n$. It is
crucial when the global geometric language is used. The importance of
this distinction is discussed in~\cite{Morales:00::}.
\begin{example}
 \textsl{
  If $\varphi(t)$ is given by rational functions of $t$ then $\Gamma$
  is the Riemann sphere $\CP$ with some points removed (poles of
  $\varphi(t)$). } 
\end{example}
\begin{example}
\textsl{
  If $\varphi(t)$ is given by elliptic functions with fundamental
  periods $T_1$ and $T_2$ then $\Gamma$ is a torus $\mathbb{T}$ with
  some points removed (poles of $\varphi(t)$). Moreover,
  $\mathbb{T}=\C/L$, where $L=\{ z\in\C \,|\,z= iT_1+jT_2, \,\,
  (i,j)\in\Z^2\}$.
}
\end{example}
Together with system \eqref{eq:ds} we also consider the 
variational equations
\begin{equation}
\label{eq:vds}
\Dt \xi = A(t) \xi, \qquad A(t) = \frac{\partial v}{\partial x}(\varphi(t)), 
\qquad \xi\in\C^n.
\end{equation}
Let us note that  one solution of the above system is known. In fact, if we 
put $\eta = v(\varphi(t))$, then 
\begin{equation}
\label{eq:dum}
\Dt \eta =  \frac{\partial v}{\partial x}(\varphi(t))\Dt \varphi(t) = 
    \frac{\partial v}{\partial x}(\varphi(t))  v(\varphi(t)) = A(t)\eta.
\end{equation}
\begin{example}
\label{ex:nves}
\textsl{
  Let us assume that system~\eqref{eq:ds} admits the following invariant set
\[
\Pi = \{ (x_1,
  \ldots ,x_n)\in\C^n\, |\, x_1=\cdots=x_{n-1}= 0 \},
\]
i.e., the right hand sides $v(x)=(v_1(x), \ldots, v_n(x))$
of~\eqref{eq:ds} are such that $v_i(x)=0$ for $i=1,\ldots n-1$ when
$x_i=0$ for $i=1,\ldots n-1$. Then a particular solution $\varphi(t)$
lies on the $n$-th coordinate axis. Obviously, we have
\[
   \frac{\partial v_i}{\partial x_n}(\varphi(t))=0,\qquad i=1,\ldots n-1.
\]
Thus, the matrix $A(t)$ has the following block form
\begin{equation}
\label{eq:At}
A(t) = \begin{bmatrix}
         B(t) & 0 \\
         b(t) & a(t)
         \end{bmatrix},
\end{equation}
where 
\[
B(t) =\left[ \frac{\partial v_i}{\partial x_j}\right],  \qquad
b(t) =\left[ \frac{\partial v_n}{\partial x_j}\right],\qquad
a(t) = \frac{\partial v_n}{\partial x_n}.
 \qquad i,j = 1, \ldots,n-1.
\]
Thus, the first $n-1$ variational equations form a
  closed sub-system of equations which are called \emph{the Normal
    Variational Equations} (NVEs).
}
\end{example}
The above example shows that the order of~\eqref{eq:vds} can be
reduced by one, at least locally. However, these local reductions can be
performed consistently over the whole $\Gamma$, so we can talk about the
NVEs associated with $\Gamma$. For a global definition of the NVEs
see~\cite{Ziglin:82::b,Braider:96::}. Here, just for simplicity, we 
assume that the coordinates $x$ are chosen  as in Example~\ref{ex:nves}.
Thus, the NVEs have the form 
\begin{equation}
\label{eq:nvds}
\Dt \tilde\xi = B(t) \tilde\xi, \qquad
 \tilde\xi=(\xi_1,\ldots, \xi_{n-1})\in\C^{n-1},
\end{equation}
where $B(t)$ is the $(n-1)\times(n-1)$ upper diagonal sub-matrix of matrix
$A(t)$, see \eqref{eq:At}.
\begin{remark}
\label{rem:nveham}
\textsl{
If system~\eqref{eq:ds} is Hamiltonian then $n$ is even ($n=2m$) and
we have one first integral, namely the Hamiltonian of the system. Then
we can reduce the order of the variational equations by two. Let for our
particular solution the value of the Hamiltonian be $E$. Then we can
restrict ~\eqref{eq:ds} to the level $H(x)=E$, and we obtain a system of
$2m-1$ autonomous equations with the same particular solution. Then we
perform the above-mentioned reduction of the corresponding variational
equations (of order $2m-1$), and we obtain the NVEs of order $2(m-1)$
which are Hamiltonian ones. The last statement follows from the
Whittaker theorem about isoenergetic reduction of order of a
Hamiltonian system.}
\end{remark}
\begin{remark}
\textsl{
  A typical situation with a Hamiltonian system is the following. For the
  investigated system with Hamiltonian function $H(x)$, $x=(q_1, p_1,
  \ldots ,q_m, p_m)\in\C^{2m}$ there exists an invariant canonical
  plane $\Pi$, e.g., 
\[
\Pi = \{ (q_1, p_1,
  \ldots ,q_m, p_m)\in\C^{2m}\, |\, q_1=p_1=\cdots =q_{m-1}=p_{m-1}= 0 \}.
\]
This implies that 
\[
   \frac{\partial H}{\partial q_i }(x)=
 \frac{\partial H}{\partial p_i }(x)=0, \qquad x\in \Pi, 
\qquad i=1, \ldots m-1.
\]
Thus, the Hessian of $H$ calculated for $x\in\Pi$ has the following
block  form
\[
H''(x) = \begin{bmatrix}
           h(x) & 0 \\
           0   & h_m(x)
       \end{bmatrix},
\]
where $h(x)$ is a symmetric $2(m-1)\times 2(m-1)$ matrix, and $h_m(x)$
is a symmetric $2\times2$ matrix. For a particular solution
$\varphi(t)\in\Pi$ the variational equations have the form
\[
 \dot \xi = J_mH''(\varphi(t)) \xi, \qquad \xi\in\C^{2m},
\]
where $J_m$ is the symplectic unit (of dimension $2m\times 2m$), and
the normal variational equations are the following
\[
 \dot{\tilde\xi} = J_{m-1}h(\varphi(t)) \tilde\xi, 
\qquad \tilde\xi\in\C^{2(m-1)}.
\]
}
\end{remark}
\begin{example}
\textsl{
  Let us consider the Hamiltonian system given by the following
  Hamiltonian function
\begin{equation}
H= \frac{1}{2}(p_1^2 + p_2^2) + \frac{1}{4}(q_1^2+q_2^4) +
  \frac{e}{2}q_1^2q_2^2,
\end{equation}
where $e$ is a parameter and $(q_1,p_1,q_2, p_2)\in\C^4$. The Hamilton's
equations for this system admit the following particular solution
$\varphi(t)=(0, 0, q_2(t), p_2(t))$, where $q_2(t) = \cn(t,k)$, $p_2(t)
= -\sn(t,k)\dn(t,k)$, $k=\sqrt{2}/2$, and $\sn$, $\cn$, $\dn$ denote
the Jacobi elliptic functions. As this particular solution lies in
the $(q_2,p_2)$ plane, the NVEs correspond to variations in $q_1$ and
$p_1$, so they have the following form 
\begin{equation}
\dot \xi = \eta, \qquad \dot\eta = - e q_2(t)^2 \xi.
\end{equation}
Note that the above system is a Hamiltonian one. It is generated by the time
dependent Hamiltonian function $h= (\eta^2 + e q_2(t)^2 \xi^2)/2$.
} 
\end{example}

In the Ziglin and Morales-Ramis theories the concepts of the monodromy
group and the differential Galois group play fundamental role.  In the
successive subsections we introduce these concepts and give
formulations of basic lemmas and theorems which we used in this paper.
\subsection{Monodromy group}

Let $\Xi(t)$ be the matrix of fundamental solutions of~\eqref{eq:vds}
defined in a neighbourhood of $t_0\in\C$, i.e., columns of $\Xi(t)$
are $n$ linear independent solutions of~\eqref{eq:vds}, and let
$\gamma$ be a closed path (with the base point at $t_0$) on the
complex time plane.  An analytic continuation of $\Xi(t)$ along
$\gamma$ gives rise to a new matrix of fundamental solutions
$\widehat\Xi(t)$ in a neighbourhood of $t_0$ which does  not necessarily
coincide with $\Xi(t)$.  However, the solutions of a linear system form
an $n$ dimensional linear space, so we have $\widehat\Xi(t) =
\Xi(t)M_\gamma$, for a certain nonsingular matrix
$M_\gamma\in\mathrm{GL}(n,\C)$ which is called the monodromy matrix.
\begin{example}
\textsl{
The system 
\[
\frac{\rmd \phantom{t}}{\rmd t} 
   \begin{bmatrix}
     \xi_1\\
     \xi_2 
    \end{bmatrix} = \frac{1}{t^2}
                    \begin{bmatrix}
                      0 & t^2 \\
                      -1 & t
                     \end{bmatrix} \begin{bmatrix}
     \xi_1\\
     \xi_2 
    \end{bmatrix}, 
\] 
has two linearly independent solutions
\[
 \xi^{(1)} = (t, 1)^T, \quad\text{and}\quad
\xi^{(2)} = (t\ln t, 1 +\ln t )^T.
\]
After continuation along a loop $\gamma$ encircling $t=0$ once,
the solution $\xi^{(1)}$ is unchanged.  However, the second solution
changes into
\[ 
(t(2\pi\rmi+\ln t), 1 +
2\pi\rmi+\ln t)^T, 
\]
and thus we have
\begin{equation*}
\label{eq:mon}
\Xi(t)=
\begin{bmatrix}
t &  t\ln t\\
1 & 1+\ln t 
\end{bmatrix} \xrightarrow[\gamma]{\qquad}
\Xi(t)M_\gamma =
\begin{bmatrix}
t &  t\ln t\\
1 & 1+ \ln t 
\end{bmatrix}
\begin{bmatrix}
1 & 2\pi i \\
0 & 1
\end{bmatrix}.
\end{equation*}
}
\end{example}
\begin{example}
\textsl{
Let us consider the following system
\[
\Dt \xi = \frac{1}{t} C\xi, 
\]
where $C$ is a constant matrix. Let $\gamma$ be a loop encircling once  
 $t=0$ counterclockwise. Then the monodromy matrix is given by
\[
M_\gamma = \exp[2\pi\rmi C].
\]
}
\end{example}
The monodromy matrix $M_\gamma$ does not depend on a particular choice of
$\gamma$.  If the path $\sigma$ can be obtained by a continuous
deformation of the path $\gamma$, then $M_\sigma=M_\gamma$.  We denote by
$[\gamma]$ the set of all paths which can be obtained by continuous
deformations of $\gamma$, and it is called the homotopy class of path
$\gamma$. Thus, the monodromy matrix $M_\gamma$ depends on the
homotopy class of path $\gamma$. If we have two paths $\sigma$ and
$\gamma$ by their product $\tau=\sigma\cdot\gamma$ we understand the path
$\tau$ obtained in the following way: first we go along $\gamma$, then
along $\sigma$. One can show that this defines properly a product of
homotopy classes, i.e.,
$[\tau]=[\sigma]\cdot[\gamma]:=[\sigma\cdot\gamma]$. We can also
define the inverse $\gamma^{-1}$ of the path $\gamma$: we go along
$\gamma$ in the opposite direction. Again we have a correct definition
$[\gamma]^{-1}:=[\gamma^{-1}]$. In this way the homotopy classes form a
group which is called the first homotopy group of a Riemann surface
(walking on the complex time plane $t$, in fact we make loops on $\Gamma$
because $t$ parametrises the surface $\Gamma$). We denote it by 
$\pi_1(\Gamma, t_0)$.
\begin{remark}
  \textsl{ If we change the base point $t_0$ of the paths, then,
    instead of the matrices $M_\gamma$, we obtain $CM_\gamma C^{-1}$,
    where $C$ is a certain nonsingular matrix (the same for all
    paths). It means that the homotopy groups at all points $t_0$ are
    isomorphic.}
\end{remark}
All the monodromy matrices form a group $\cM$ with respect to matrix
multiplication which is a subgroup of $\mathrm{GL}(n,\C)$. From the
definition of monodromy we have $M_{\sigma\cdot\gamma}=
M_{\gamma}M_{\sigma}$, so $M_{[\sigma\cdot\gamma]}=M_{[\gamma]}
M_{[\sigma]}$. In the same way $M_{[\gamma]^{-1}}= M_{[\gamma]}^{-1}$.
In other words, the monodromy matrices form an anti-representation of
$\pi_1(\Gamma, t_0)$ in $\mathrm{GL}(n,\C)$.
\begin{remark}
\textsl{
  If system~\eqref{eq:ds} is Hamiltonian, then the variational
  system~\eqref{eq:vds} is also a Hamiltonian one, and the monodromy
  group is a subgroup of the symplectic group $\mathrm{Sp}(2m,\C)$ where
  $2m=n$. If we consider the NVEs for a Hamiltonian system as it was
  described in Remark~\ref{rem:nveham}, then the monodromy group of
  these equations is contained in $\mathrm{Sp}(2(m-1),\C)$. }
\end{remark} 
\subsection{Basic lemma of the Ziglin theory}
Let us assume that $F(x)$ is a holomorphic first integral
of~\eqref{eq:ds}. The Taylor expansion of $F(\varphi(t)+\xi)$ has the
form
\begin{equation}
\label{eq:Ftay}
F(\varphi(t)+\xi) = F(\varphi(t)) + F_m(t,\xi) + \cdots,
\end{equation}
where $ F_m(t,\xi)$ is a homogeneous polynomial (with respect to
the coordinates of $\xi$) of degree $m>0$. It is easy to show that $
F_m(t,\xi)$ is a first integral of the variational
equations~\eqref{eq:vds}. We called $ F_m(t,\xi)$ the leading term of
the first integral. When the first integral $F(x)$ is a meromorphic
function, then it can be represented as a ratio $P(x)/Q(x)$ of two
holomorphic functions $P(x)$ and $Q(x)$. If $P_m(t,\xi)$ is the
leading term of $P(x)$ and $Q_k(t,\xi)$ is the leading term of $Q(x)$,
then by the leading term of $F(x)$ we understand $P_m(t,\xi)/Q_k(t,\xi)$,
and it is a first integral of~equations~\eqref{eq:vds} which is
rational with respect to $\xi$.

An analytic continuation of solutions of~\eqref{eq:ds} along a closed
path $\gamma$ transforms initial conditions for these solutions to
other points in the following way. At $t_0$ we start from $\xi_0$.
For small $t$ we move along $\gamma$ and $\xi_0$ goes to
$\xi(t)=\Xi(t)\xi_0$. After continuation, we return to a neighbourhood
of $t_0$, but now our point is moved to $\widehat\Xi(t)\xi_0$, and
thus at the end of the path at $t_0$ we obtain the point
\[
\widehat\Xi(t_0)\xi_0=
\Xi(t_0)M_\gamma\xi_0=M_\gamma\xi_0,
\]
as $\Xi(t_0)$ is the identity.  Thus we have the following map
\[
 (t_0,\xi_0) \xrightarrow[\gamma]{\qquad} (t_0,M_\gamma\xi_0). 
\]
It is important to notice here that $t_0$, as well as $\xi_0$, are
\emph{arbitrary}.

Let $F_m(t,\xi)$ be a first integral of~\eqref{eq:vds} and let
$F_m^0=F_m(t_0,\xi_0)$. A first integral does not change its value
when we make an analytic continuation. Thus taking the loop $\gamma$ we
have
\[
F_m^0=F_m(t_0,\xi_0) = F_m(t_0, M_\gamma \xi_0).
\]
As $t_0$, $\xi_0$ and $\gamma$ are arbitrary we have 
\begin{equation}
\label{eq:Fminv}
F_m(t,\xi) = F_m(t, M_\gamma \xi),
\end{equation}
for all $M_\gamma\in\cM$. In other words, $F_m(t,\xi)$ is invariant
with respect to the natural action of the monodromy group. A
non-constant function satisfying the above condition is called a first
integral (or an invariant) of the monodromy group (polynomial
(rational) if $F_m$ is a polynomial (rational) function of the coordinates
of $\xi$). We can repeat all the above considerations for the normal
variational equations.
The condition~\eqref{eq:Fminv} is restrictive. When the monodromy group of
the NVEs is `big', then it can happen that there is no non-constant
polynomial (rational) invariant, and this fact implies that
system~\eqref{eq:ds} does not have a holomorphic (meromorphic) first
integral.

The following lemma formulated by Ziglin gives the necessary condition
for integrability, see Proposition on p.~183 in \cite{Ziglin:82::b}
and Proposition on p.~4 in \cite{Ziglin:97::}.
\begin{lemma}
\label{lem:zig}
If system \eqref{eq:ds} possesses a meromorphic first integral defined
in a neighbourhood $U\subset M^n$, such that the fundamental group of
$\Gamma$ is generated by loops lying in $U$, then the monodromy group
$\cM$ of the normal variational equations  has a
rational first integral.
\end{lemma}
\begin{remark}
  \textsl{ The reason why in the above Lemma the necessary condition
    for integrability cannot be formulated (or, rather, it is more
    difficult to formulate) in terms of the monodromy group of the full
    variational equations is the following. The monodromy group
    of~\eqref{eq:vds} always possesses one polynomial invariant.  Let
    us explain why.  As it was mentioned, for equations~\eqref{eq:vds}
    we know one particular solution $\eta = v(\varphi(t))$,
    see~\eqref{eq:dum}. If $\Xi(t)$ is the fundamental matrix of
    \eqref{eq:vds}, then we can find a vector $c\in\C^n$ such that
    $\eta = \eta(t)=\Xi(t)c$. Let us assume for simplicity that
    the solution $\varphi(t)$ is single-valued. Thus the continuation of $
    \eta(t)$ along an arbitrary path $\gamma$ does not change it, and
    we have that $\eta(t)= \widehat\Xi(t)c =\Xi(t)M_\gamma c =
    \Xi(t)c$. It follows that $M_\gamma c = c$, i.e., the vector $c$ is an
    eigenvector of all monodromy matrices and it corresponds to an 
    eigenvalue 1. Thus, in appropriate coordinates
    $x=(x_1,\ldots,x_n)$, the monodromy matrices $M$ can be put
    simultaneously into the following form
\[
  M = \begin{bmatrix}
          1 & 0 \\
          m & \widetilde M
       \end{bmatrix},
\]
where $m$, $\widetilde M$ are $(n-1)\times 1$ and $(n-1)\times (n- 1)$
matrices, respectively.  But now the linear polynomial $f(x)=x_1$ is an
invariant of the monodromy group.  }
\end{remark}
\subsection{Differential Galois group}
Let us assume that the entries of the matrix $A(t)$ of the linear
system~\eqref{eq:vds} are rational functions of $t$. We know that
solutions of linear equations with rational coefficients are not
necessarily rational, however, we can ask whether a given linear
equation or a system of linear equations is solvable in terms of
`known' functions.  This question was investigated at the end of the
nineteenth and at the beginning of the twentieth century by Picard,
Vessiot and others. Later on, thank to works of Kolchin, the
Picard-Vessiot theory was considerably developed to what is now called
the differential Galois theory.  For a general introduction to this
theory see \cite{Singer:90::,Kaplansky:76::,Beukers:92::,Magid:94::}.

Through this subsection our leading example is a linear  second
order differential equation with rational coefficients
\begin{equation}
\label{eq:gle}
w'' + p w' + q w = 0, \qquad p, q\in\C(t),  
  \qquad '\equiv \frac{\rmd\phantom{t}}{\rmd t}.
\end{equation}
In what follows we keep algebraic notation, e.g., by $\C[t]$ we denote
the ring of polynomials of one variable $t$, $\C(t)$ is the field of
rational functions, etc. Here we consider the field $\C(t)$ as a
differential field, i.e., a field with distinguished differentiation.
Note that in our case all elements $a\in\C(t)$ such that $a'=0$ are
just constant, i.e., we have $a'=0 \Leftrightarrow a\in\C$. Thus such
elements form a field---the field of constants.
\begin{remark}
  \textsl{ In the most general case we meet in applications,
    the coefficients of~\eqref{eq:vds} are meromorphic functions defined
    on a Riemann surface $\Gamma$, which is usually denoted by 
    $\cM(\Gamma)$. Meromorphic functions on $\Gamma$ form a field. It
    is a differential field if equipped with  ordinary
    differentiation.  }
\end{remark}
The field $\C(t)$  can be extended  to a larger differential
field $K$ such that it will contain all solutions of
equation~\eqref{eq:gle}.  The smallest differential field $K$
containing $n$ linearly independent solutions of~\eqref{eq:vds} is
called the Picard-Vessiot extension of $\C(t)$ (additionally we need
the field of constants of $K$ to be $\C$).
\begin{remark}
\textsl{
  The Picard-Vessiot extension for equation~\eqref{eq:gle} can be
  constructed in the following way. We take two linearly independent
  solutions $\xi$ and $\eta$ of~\eqref{eq:gle} (we know that such
  solutions exist).  Then, as $K$ we take all rational functions of
  five variables $(t, \xi,\xi',\eta,\eta')$, i.e.,
  $K=\C(t,\xi,\xi',\eta,\eta')$.
}
\end{remark} 
\begin{remark}
\textsl{
  In the case considered (a system of complex linear equations with rational
  coefficients) the existence of the Picard-Vessiot extension follows
  from the Cauchy existence theorem. In abstract settings, i.e. when
  we consider a differential equation with coefficients in an abstract
  differential field, the existence of the Picard-Vessiot extension is
  a non-trivial fact, see e.g. \cite{Magid:94::}.
}
\end{remark}
 
Now, it is necessary to define what we understand by `known'
functions.  Informally, these are rational and algebraic functions,
their integrals  and exponential of their integrals.  More precisely, we
say that a solution $\eta$ of~\eqref{eq:gle} is:
\begin{enumerate}
\item \emph{algebraic} over $\C(t)$ if $\eta$ satisfies a polynomial
  equation with coefficients in $\C(t)$,
\item \emph{primitive} over $\C(t)$ if $\eta'\in\C(t)$, i.e., if
  $\eta=\int a$, for certain $a\in\C(t)$,
\item \emph{exponential}  over $\C(t)$ if $\eta'/\eta\in\C(t)$, i.e., if
  $\eta=\exp\int a$, for certain $a\in\C(t)$.
\end{enumerate}
We say that a differential field $L$ is a Liouvillian extension of $\C(t)$
if it can be obtained by successive extensions
\[
\C(t)=K_0\subset K_1 \subset \cdots  \subset K_m = L,
\]
such that $K_i=K_{i-1}(\eta_i)$ with $\eta_i$ either algebraic,
primitive or exponential over $K_ {i-1}$. Our vague notion 'known'
functions means Liouvillian functions.  We say that~\eqref{eq:vds} is
solvable if for it the Picard-Vessiot extension is a Liouvillian
extension.
\begin{remark}
\textsl{
  All elementary functions, like $\rme^t$, $\log t$, trigonometric
  functions, are Liouvillian, but special functions like Bessel or Airy
  functions are not Liouvillian.
}
\end{remark}
\begin{example}
\textsl{
The equation
\[
4t w'' + 2 w' - w = 0,
\]
has two linearly independent solutions $w_1= exp[ \sqrt{t}]$ and $w_2=
exp[-\sqrt{t}]$. Both of them are Liouvillian. 
}
\end{example}

How can we check if solutions of a given equation are Liouvillian? For
this purpose we need to check properties of the differential Galois
group of the equation. This group can be defined as follows. For the
Picard-Vessiot extension $K\supset\C(t)$ we consider all automorphisms
of $K$ (i.e. invertible transformations of $K$ preserving field
operations) which commute with differentiation. An automorphism
$g:K\to K$ commutes with differentiation if $g(a')=(g(a))'$,
for all $a\in K$.  We denote by $\cA$
the set of all such automorphisms.  Let us note that automorphisms
$\cA$ form a group. The differential Galois group $\cG$ of extension
$K\supset\C(t)$, is, by definition, a subgroup of $\cA$ such that it
contains all automorphisms $g$ which do not change elements of
$\C(t)$, i.e., for $g\in\cG$ we have $g(a)=a$ for all $a\in\C(t)$.
\begin{remark}
\textsl{
  It seems that the definition of the differential Galois group is
  abstract and that it is difficult to work with it. However, from this
  definition we can deduce that it can be considered  as a subgroup of
  invertible matrices. Let $\cG$ be the differential Galois group of
  equation~\eqref{eq:gle} and let $g\in\cG$. Then we have
\[
0=g(0)=g(w'' + p w' + q w)= g(w'')+g(p)g(w') +g(q)g(w),
\]
but $g$ commutes with differentiation so $g(w'')=(g(w))''$,
$g(w')=(g(w))'$, and, moreover, $g(p)=p$, $g(q)=q$ because
$p,q\in\C(t)$. Thus we have
\[
(g(w))'' +p (g(w))' + q g(w)=0.
\]
In other words, if $w$ is a solution of equation~\eqref{eq:gle} then
$g(w)$ is also its solution. Thus, if $\xi$ and $\eta$ are linearly
independent solutions of~\eqref{eq:gle}, then
\[
   g(\xi) = g_{11} \xi + g_{21} \eta, \qquad
   g(\eta)  = g_{12} \xi + g_{22} \eta, 
\] 
and 
\[
  g\left( \begin{bmatrix}
    \xi & \eta\\
    \xi' & \eta'
   \end{bmatrix}\right) = 
 \begin{bmatrix}
    \xi & \eta\\
    \xi' & \eta'
   \end{bmatrix}
 \begin{bmatrix}
    g_{11} & g_{12} \\
    g_{21} & g_{22}
   \end{bmatrix}.
\]
Hence, we can associate with an element $g$ of the differential Galois
group $\cG$ an invertible matrix $[g_{ij}]$, and thus we can consider
$\cG$ a subgroup of $\mathrm{GL}(2,\C)$.  If instead of the solutions $\xi$
and $\eta$ we take other two linearly independent solutions, then all
matrices $[g_{ij}]$ are changed by the same similarity transformation.
}
\end{remark}
The construction presented in the above remark can be easily generalised to a
linear differential equation of an arbitrary order and to a system of
linear equations.  Thus we can treat the differential Galois group as
a subgroup of $\mathrm{GL}(n,\C)$. Let us list basic facts about the
differential Galois group
\begin{enumerate}
\item If $g(a)=a$ for all $g\in\cG$, then $a\in\C(t)$. 
\item Group $\cG$ is an algebraic subgroup of $\mathrm{GL}(n,\C)$.
  Thus it has a unique connected component $\cG^0$ which contains the
  identity, and which is a normal subgroup of finite index.
\item Every solution of the differential equation is Liouvillian if
  and only if $\cG^0$ conjugates to a subgroup of the triangular group.
  This is the Lie-Kolchin theorem.
\end{enumerate}
For proofs and details we refer the reader to the cited references.
\subsection{Basic theorem of the Morales-Ramis theory}
For a given linear system of linear differential equations we can
determine the  monodromy group $\cM$ and the differential Galois group
$\cG$. From the description given above it follows that both these
groups are related. In fact, we have $\cM \subset\cG$.  In other words,
the differential Galois group $\cG$ is `bigger' then the monodromy
group $\cM$.
\begin{example}
  \textsl{ For the Airy equation $\ddot x = tx$ the monodromy group is
    trivial, i.e., it contains only one element---the identity matrix,
    while its differential Galois group is $\mathrm{SL}(2,\C)$. For a
    proof see e.g. \cite{Kaplansky:76::}. }
\end{example}
\begin{remark}
  \textsl{ It should be mentioned that the determination of the
    monodromy group is a difficult task, and this groups is known only
    for a very limited number of equations. What concerns the
    determination of the differential Galois group we are in much
    better situation. There exist algorithms (the Kovacic algorithm
    \cite{Kovacic:86::}) which allow to determine this group for an
    arbitrary second order linear differential equation with rational
    coefficients (see the Appendix for additional references). }
\end{remark}
The fact that $\cM \subset\cG$ suggests  the use of $\cG$ instead of
$\cM$ to formulate a necessary condition for non-integrability.  If
system \eqref{eq:ds} possesses a meromorphic first integral, then
\eqref{eq:vds} has a first integral and this fact imposes a
restriction on its differential Galois group $\cG$, as it imposes
restrictions on its monodromy group $\cM$. In fact, we have a lemma
which is analogous to Lemma~\ref{lem:zig}.
\begin{lemma}
\label{lem:mor}
If system \eqref{eq:ds} possesses a meromorphic first integral defined
in a neighbourhood $U\subset M$ of $\Gamma$, then the differential
Galois group $\cG$ of the NVEs has a rational first integral.
\end{lemma}
The above lemma is a variant of Lemma~III.1.13 from \cite{Audin:01::},
see also Lemma~4.6 in \cite{Morales:99::}. For proof and details see
Chapter III of \cite{Audin:01::}. 

The differential Galois theory gives a powerful tool to the study of 
integrability of Hamiltonian systems. The Morales-Ramis theory is
formulated in the most exhaustive form in book \cite{Morales:99::} and papers 
\cite{Morales:01::b}.
It gives a necessary condition of integrability of a Hamiltonian
system for which we know a non-equilibrium solution.  The main theorem
is the following
\begin{theorem}
\label{th:MR}
Assume that the Hamiltonian system is integrable in the Liouville
sense in a neighbourhood of a particular solution. Then the identity
component of the differential Galois group of the NVEs is Abelian.
\end{theorem}

%
%
\section{Particular solutions and variational equations}
\label{sec:var}
From now on we consider~\eqref{eq:bel} as a complex system, i.e., we
assume that $(\bM,\bN,\bS)\in\C^9$ and $t\in\C$.  Without loss of
generality, choosing appropriately the unit of time and length, we can
put $\omega_\mathrm{K}=\omega_\mathrm{O}=1$ and $A=1$. According to
our knowledge, for an arbitrary $\bL$, equations~\eqref{eq:bel} do
not admit a particular solution. However, if we assume that $\bL$
coincides with one of the principal axes, e.g., $\bL=[0,0,1]^T$, then
one can find particular solutions.  In fact, in this case the
following manifold
\begin{equation}
\label{eq:inv}
\cN=\{(\bM,\bN,\bS)\in\C^9\,|\,\,
M_2=M_3=N_2=N_3=S_1=0,\,\,\,N_1=1\},
\end{equation}
is invariant with respect to the flow generated by
system~\eqref{eq:bel}.  Solutions lying on $\cN$ describe the planar
rotations of the satellite when its third axis is permanently in the
orbital plane and its first axis is perpendicular to the orbital
plane. Moreover, we can easily find an analytic form of the solutions
of~\eqref{eq:bel} describing this motion.  In fact,
system~\eqref{eq:bel} restricted to $\cN$ has the form
\begin{equation}
\label{eq:rest}
\dot{M}_1=3(C-B)S_2S_3,\qquad
\dot{S}_2=(\Omega_1-1)S_3,\qquad
\dot{S}_3=-(\Omega_1-1)S_2,
\end{equation}
and it possesses two first integrals
\begin{equation}
\label{eq:hrest}
H_{|\cN}=\dfrac{1}{2}M_1^2-M_1+\dfrac{3}{2}\left(BS_2^2+CS_3^2\right),
\qquad 
H_{2|\cN}=S_2^2+S_3^2.
\end{equation}
We can introduce on the level $H_{2|\mathcal{N}}=1$ local coordinate
$\phi$ such that
\[
S_2=-\cos\phi\quad\text{and}\quad S_3=\sin\phi.
\]
Then system \eqref{eq:rest} reads
\begin{equation}
\label{eq:firest}
\dot{M}_1=-3(C-B)\sin\phi\cos\phi,\qquad
\dot{\phi}=M_1-1.
\end{equation}
Thus, we have 
\begin{equation}
\label{eq:vfi}
\ddot \varphi = -3(C-B)\sin \varphi, \qquad \varphi =2 \phi.
\end{equation}
Solving the above equation we obtain an one parameter family $\Phi(t,k)$
of the solutions of~\eqref{eq:bel} expressed in terms of the Jacobi
elliptic functions. Let us define
\begin{equation}
\label{eq:omega}
\omega=\sqrt{3|C-B|}.
\end{equation}
Then the explicit form of the solutions is given by 
\begin{equation}
\label{eq:M1t}
M_1(t,k) =1+\omega k\cn (\omega t,k),
\end{equation}
and for $C>B$
\begin{equation}
\label{eq:sCgB}
\begin{split}
S_2(t,k)&=-\dn (\omega t,k),\\
S_3(t,k)&= k\sn (\omega t,k);
\end{split}
\end{equation} 
for $C<B$ we have
\begin{equation}
\label{eq:sBlC}
\begin{split}
S_2(t,k)&=k\sn (\omega t,k),\\
S_3(t,k)&=\dn (\omega t,k),
\end{split}
\end{equation} 
where
\begin{equation}
k=\sqrt{\dfrac{\omega^2+E}{2\omega^2}}\in(0,1),
\end{equation}
and $E$ is the value of the energy integral for equation
\eqref{eq:vfi}, i.e.,
\[
E = \frac{1}{2}{\dot\varphi}^2 -\omega^2\cos\varphi.
\]
Let us note that for the above solutions we have
\begin{equation}
\label{eq:HPhi}
H(\Phi(t,k)) = \frac{1}{2}\omega^2k^2 + \frac{1}{2}(3B-1):=h(k).
\end{equation}
From the above formulae it follows that the  particular solutions given above 
are single-valued, meromorphic, and double periodic with periods
\[
  T(k)= \frac{4}{\omega}K(k),\qquad 
T'(k)= \frac{4}{\omega}\rmi K'(k),
\]
where $K(k)$ is the complete elliptic integral of the first kind with
modulus $k$, $K'(k):=K(k')$, and $k':=\sqrt{1-k^2}$. In each period
cell they have four simple poles at:
\begin{equation}
\label{eq:si}
\begin{split}
 \tau_1(k)&= \frac{1}{2}T(k)+ \frac{1}{4}T'(k), \qquad 
 \tau_2(k)=  \tau_1(k) + \frac{1}{2}T'(k), \\
\tau_3(k)&=  \tau_2(k)  + \frac{1}{2}T(k), \qquad 
\tau_4(k)= \tau_3(k) - \frac{1}{2}T'(k) \quad \mod (T(k),T'(k)).
\end{split}
\end{equation}
Thus, the Riemann surfaces $\Gamma_k$ associated with the particular
solutions $\Phi(t,k)$ are tori with four points: $s_l(k)=
\Phi(\tau_l(k),k)$, $l=1,2,3,4$ removed.  In $\C^3$ with coordinates
$(M_1,S_2,S_3)$ these Riemann surfaces are intersections of two
quadrics
\begin{equation}
\label{eq:quadrics}
\dfrac{1}{2}M_1^2-M_1+\dfrac{3}{2}\left(BS_2^2+CS_3^2\right)=h(k), 
\qquad S_2^2+S_3^2=1.
\end{equation}   
For $0<k<1$ the four points $s_l(k)$ correspond to four points of
intersections of the above quadrics at infinity. 

As our aim is to investigate the case when the satellite is symmetric, we
assume that $A=B=1$.  For a symmetric satellite, we have one more
first integral, namely $H_5 = M_3$.  This first integral is connected
with the existence of an one parameter symmetry of the system.  Equations
\eqref{eq:bel} are invariant (for the prescribed choice of $\bL$ and
the symmetry axis) with respect to an action of group
$\mathrm{SO}(2,\R)$. Simply, the principal axes perpendicular to the
symmetry axis of the body can be chosen arbitrarily. Thanks to that, we
can reduce the number of degrees of freedom by one. Thus, the reduced
system is  Hamiltonian  with two degrees of freedom and it depends
parametrically on the value of the chosen level of $H_5$.

Further calculations can be performed in  the $(\bM,\bN,\bS)$ coordinates
in the same way as it was done in \cite{Audin:02::}. Here we perform
them in  canonical coordinates on $\cM^6$.  This approach allows to
deduce the normal variational equations in an elementary way.
Appropriate canonical variables on $\cM^6$ can be chosen in the
following way.  We parametrise the orientation of the principal axes
of the body with respect to the orbital reference frame by the Euler
angles  $ (q_1,q_2,q_3)$ of the type 3-1-3, and we take them as
generalised coordinates. Then generalised momenta conjugated to $
(q_1,q_2,q_3)$ are given by
\begin{equation}
\label{eq:pM}
\bp = \bK\bM, \qquad 
\bK= \begin{bmatrix}
       \sin q_3 \sin q_2 & \cos q_3 \sin q_2& \cos q_2 \\
       \cos q_3          & -\sin q_3        &    0 \\
        0                &                0 &    1
     \end{bmatrix}.
\end{equation}
Moreover, we have
\begin{equation}
\label{eq:NS}
\bN = \begin{bmatrix}
 \sin q_3 \sin q_2\\ \cos q_3 \sin q_2\\ \cos q_2
 \end{bmatrix} \qquad
\bS = \begin{bmatrix}
    - \sin q_3 \cos q_2 \sin q_1 + \cos q_3 \cos q_1 \\
   -\cos q_3 \cos q_2 \sin q_1 - \sin q_3 \cos q_1 \\
   \sin q_2 \sin q_1
\end{bmatrix}.
\end{equation}
In the introduced canonical coordinates the Hamiltonian~\eqref{eq:ener}
reads
\begin{equation}
\label{eq:Hcan}
\begin{split}
H = &\frac{1}{2}\left(\frac{p_3 \cos q_2 - p_1}{\sin q_2}\right)^2 + 
  \frac{1}{2}p_2^2 + \frac{1}{2C}p_3^2-p_1 + \\
  & +   \frac{3}{2}(C-1) \sin^2q_1\sin^2 q_2 -
\frac{1}{2}\xi \cos^2 q_2.
\end{split}
\end{equation} 
As we can see, $q_3$ is a cyclic coordinate and $p_3=M_3$ is a first
integral. Thus, considering $p_3$ as an additional parameter, $H$
defines a Hamiltonian system with two degrees of freedom with
$x=(q_1,q_2,p_1,p_2)$ as canonical coordinates. As our particular
solutions lie on the level $M_3=0$, we investigate this system for
$p_3=0$ , i.e, we consider the Hamiltonian system given by the
following Hamiltonian
\begin{equation}
\label{eq:Hp30}
H = \frac{1}{2}\frac{p_1^2}{\sin^2 q_2} + 
  \frac{1}{2}p_2^2 - p_1
   +   \frac{3}{2}(C-1) \sin^2q_1\sin^2 q_2 -
\frac{1}{2}\xi \cos^2 q_2.
\end{equation} 
Now, the invariant manifold $\cN$ corresponds to the canonical plane
$q_2=\pi/2$, $ p_2=0$, on which canonical equations generated by $H$
have the form
\begin{equation}
\label{eq:q2p20}
\dot q_1 = p_1 - 1, \qquad  \dot p_1 = -3(C-1)\sin q_1 \cos q_1. 
\end{equation}
Comparing them with equations~\eqref{eq:firest} we see that $p_1= M_1$
and $q_1 = \phi$ (note that this fact follows from the definition of
$\cN$, formulae~\eqref{eq:pM}, \eqref{eq:NS} and the fact that on
$\cN$ we have $q_3 = \pi/2$). Thus, the explicit form of the particular
solutions $x=x(t,k)=(q_1(t,k),\pi/2, p_1(t,k),0)$ is given by
\begin{equation}
\label{eq:p1tk}
p_1(t,k) =1+\omega k\cn (\omega t,k),
\end{equation}
and for $C>1$
\begin{equation}
\label{eq:sCg1}
\begin{split}
\cos q_1(t,k)&=\dn (\omega t,k),\\
\sin q_1 (t,k)&=k\sn (\omega t,k),
\end{split}
\end{equation} 
and for $C<1$
\begin{equation}
\label{eq:sBl1}
\begin{split}
\cos q_1(t,k)&=-k\sn (\omega t,k),\\
\sin q_1(t,k)&=\dn (\omega t,k).
\end{split}
\end{equation} 
We note here that for a symmetric satellite we have 
\[
\omega=\sqrt{3|C-1|}\in(0,\sqrt 3).
\]
The variational equations along the particular solution $x(t,k)$ have the
following form
\begin{gather}
\dot Q_1 = P_1, \qquad \dot P_1 = 3(1-C)\cos\,(2q_1(t,k)) Q_1 , \\ 
\label{eq:Q2P2}
\dot Q_2 = P_2 , \qquad \dot P_2 =[\xi -p_1(t,k)^2+3(C-1)\sin^2q_1(t,k)]Q_2.
\end{gather}
As the particular solutions lie in the plane $\{q_2=\pi/2, p_2=0\}$,  the NVEs
correspond to the subsystem~\eqref{eq:Q2P2} which can be written as a
second order linear equation of the form
\begin{equation}
\label{eq:nvet}
\ddot Q + a(t,k)Q=0, \qquad Q\equiv Q_2,
\end{equation}
where
 \begin{equation}
 \label{eq:avar}
  a(t,k) = \begin{cases}
            (1+k\omega\cn(\omega t,k))^2+\omega^2\dn^2(\omega
            t,k)-\xi\quad\text{for}\quad C <1, 
           \\[\medskipamount]
            (1+k\omega\cn(\omega t,k))^2-\omega^2k^2\sn^2(\omega t,k)
            -\xi\quad\text{for}\quad C >1.  
           \end{cases}
\end{equation}
\begin{remark}
  \textsl{ Let us notice that for equation~\eqref{eq:nvet} the
    differential Galois group is a subgroup of $\mathrm{SL}(2,\C)$. It
    is always the case when a second order linear differential
    equation does not contain a term proportional to the first
    derivative.  }
\end{remark}
\begin{remark}
  \textsl{ Here we underline that the obtained  NVE~\eqref{eq:nvet} is the
    reduced normal variational equation for~\eqref{eq:bel} when
    $A=B=1$ and $\bL=(0,0,1)$ derived for
    solution~\eqref{eq:M1t}--\eqref{eq:sBlC}.  We just performed a
    symplectic reduction as in~\cite{Audin:02::}, but for this purpose
    we use local canonical coordinates.  }
\end{remark}

Equation~\eqref{eq:nvet} is defined on $\Gamma_k$. In order to use the
differential Galois theory efficiently, it is crucial to transform the
investigated equation into an equation with rational coefficients. In
our case we can do this making the following transformation
\begin{equation}
\label{eq:ttoz}
 t \longrightarrow  z := k\cn (\omega t,k).
\end{equation}
Then the NVE has the form  
\begin{equation}
\label{eq:zvar}
Q'' + p(z) Q' + q(z) Q = 0, \qquad '\equiv \frac{\mathrm{d}}{\mathrm{d} z}, 
\end{equation}
where
\begin{equation}
\begin{split}
\label{eq:p}
p(z)&= \dfrac{z\left( -1 + 2\left( k^2 - z^2 \right)  \right) }
  {\left( k^2 - z^2 \right) \left( z^2 + {k'}^2 \right) },\\
q(z)&=\begin{cases}
\dfrac{-\xi + {\left( 1 + \omega z \right) }^2 + 
    \omega^2\left( z^2 + {k'}^2 \right) }{\omega^2\,
    \left( k^2 - z^2 \right)\left( z^2 + {k'}^2 \right) },
\qquad\text{for}\,\,\,
C<1,\\
\dfrac{-\xi + {\left( 1 + \omega z \right) }^2 - 
    \omega^2\left( k^2 - z^2 \right) }{\omega^2
    \left( k^2 - z^2 \right) \left( z^2 + {k'}^2 \right) },
\qquad\text{for}\,\,\,
C>1.\end{cases}
\end{split}
\end{equation}
Equation \eqref{eq:zvar} is Fuchsian (see Appendix) and it has five
regular singular points over $\CP$, namely $z_{1,2}=\pm k$,
$z_{3,4}=\pm \rmi k'$ and $z_5=\infty$.
\begin{remark}
\textsl{
Our transformation~\eqref{eq:ttoz} is a double covering
\[
  \CP\longrightarrow\C\longrightarrow  \Gamma_k. 
\]
The differential Galois groups of equation~\eqref{eq:nvet} and
equation~\eqref{eq:zvar} are different, however these groups have the
same identity components, see~\cite{Morales:99::}.
}
\end{remark}
Changing the dependent variable 
\begin{equation}
\label{eq:tran}
   Q = W \exp\left[ -\frac{1}{2} \int_{z_0}^z p(s)\, ds \right],
\end{equation}
we   transform \eqref{eq:zvar} to the reduced form
\begin{equation}
\label{eq:normal}
  W'' = r(z) W, \qquad r(z) = -q(z) + \frac{1}{2}p'(z)  + \frac{1}{4}p(z)^2.
\end{equation}
The rational  coefficient $r(z)$ has the following  simple fraction expansion 
\begin{equation}
\label{eq:sf}
  r(z) = \sum_{k=1}^4 \left[ \frac{a_i}{ (z -z_i)^2 } +  
    \frac{b_i}{ z -z_i }\right],
\end{equation}
with coefficients
\[
a_1 = a_2 = a_3=a_4=-\dfrac{3}{16}, 
\]
and for $C<1$
\begin{gather}
 b_1= \dfrac{3\omega^2(3+4k^2)+8(1-\xi+2k\omega)}{16k\omega^2},\qquad
b_2=-b_1 +\frac{2}{\omega}, \\   
b_3=\rmi\dfrac{\omega^2(12{k'}^2+1)+8(\xi-1+2\rmi k'\omega)}{16k'\omega^2},
\qquad
b_4=b_3^*,
\end{gather}
where ${}^*$ denotes the complex conjugation.
For $C>1$ the coefficients $b_i$ are the following
\begin{gather} 
 b_1= \dfrac{\omega^2(12k^2+1)+8(1-\xi+2k\omega)}{16k\omega^2}, \qquad  
 b_2=-b_1 +\frac{2}{\omega}, \\ 
b_3=\rmi\dfrac{3\omega^2(3+4{k'}^2)+8(\xi-1+2\rmi k'\omega)}{16k'\omega^2}, 
            \qquad b_4=b_3^*.
\label{coeff}        
\end{gather}
The Laurent expansion of $r(z)$ at infinity in both cases has the same form
\begin{equation}
r(z)=\dfrac{2}{z^2}+O\left(\dfrac{1}{z^3}\right).
\label{nies}
\end{equation}
\begin{remark}
  \textsl{ Transformation~\eqref{eq:tran} changes the
    differential Galois group. For equation~\eqref{eq:normal} $\cG$ is
    a subgroup of $\mathrm{SL}(2,\C)$ but for equation~\eqref{eq:zvar}
    $\cG$ is not a subgroup of $\mathrm{SL}(2,\C)$. Generally, when the 
    coefficients $p(z)$ and $q(z)$ in ~\eqref{eq:zvar} are arbitrary
    rational functions, transformation~\eqref{eq:tran} changes also
    the identity component of $\cG$, e.g. $\cG^0$ of
    equation~\eqref{eq:zvar} can be non-Abelian but for the
    transformed equation~\eqref{eq:normal} $\cG$ can be Abelian.
    However, if $\cG^0$ of equation~\eqref{eq:zvar} is solvable then
    $\cG^0$ of equation~\eqref{eq:normal} has the same property. In
    our case transformation~\eqref{eq:tran} has the following form
\[ 
W= [(k^2-z^2)(k'^2+ z^2)]^{1/4}Q
\]  
and thus it does not change the identity component of the differential
Galois group of equation~\eqref{eq:zvar}.  This is not accidental. In
the time parametrisation the NVE has the form~\eqref{eq:nvet} and its
differential Galois group is contained in $\mathrm{SL}(2,\C)$. Then we
make transformation~\eqref{eq:ttoz} which is a finite covering, and
thus it does not change the identity component of the differential
Galois group, see Proposition~4.7 in \cite{Braider:96::}.  Then, by Lemma~4.24 from~\cite{Braider:96::} transformation~\eqref{eq:tran} has the form $W=RQ$,
where $R^n$ is a rational function for an integer $n$.}
\end{remark}
\section{Complex non-integrability}
\label{sec:non}
First, we investigate local monodromy of equation~\eqref{eq:normal} at
infinity. In many cases it simplifies proofs considerably. 
\begin{lemma}
\label{lem:log}
Let us assume that $C\neq 1$ and $2\xi\neq 3(1-C)$.
Then the local monodromy of equation~\eqref{eq:normal} at infinity is 
\begin{equation*}
M_\infty = \begin{bmatrix}
              1 & 2\pi \rmi \\
              0 & 1
             \end{bmatrix}.
\end{equation*}
\end{lemma}
\begin{proof}
  We prove the lemma for $C<1$. For $C>1$ the proof is similar. First
  we change the dependent variable $z= 1/\zeta$. This change moves
  $z=\infty$ to $\zeta=0$ and transforms~\eqref{eq:normal} to the form
\begin{equation}
\label{eq:wuzet}
W'' +\frac{2}{\zeta}W' -\frac{1}{\zeta^4}r\left(\frac{1}{\zeta}\right)W=0.
\end{equation} 
Moreover, we have 
\begin{equation}
\frac{1}{\zeta^4}r\left(\frac{1}{\zeta}\right) = \frac{2}{\zeta^2} + 
O(\zeta^{-1}),
\end{equation}
and thus the indicial equation (see~\cite[Chapter X]{Whittaker:65::})
reads
\begin{equation}
\rho(\rho-1) +2\rho - 2=0. 
\end{equation}
Hence, exponents at $\zeta=0$ are $\rho_-=-2$ and $\rho_+ = 1$. Their
difference $m = \rho_+-\rho_-=3$ is an integer, and thus, in a
neighbourhood of $\zeta=0$ one solution of~\eqref{eq:wuzet} has the
form
\begin{equation}
\label{eq:w1}
W_1(\zeta) = \zeta^{\rho_+} f(\zeta), \qquad
 f(\zeta) = 1 + \sum_{k=}^\infty f_{k}\zeta^k,
 \end{equation}
 where the series defining $f(\zeta)$is  convergent in the considered region
 \cite{Whittaker:65::}. The second solution, independent of
 $W_1(\zeta)$, is defined by the integral
\begin{equation}
\label{eq:w2}
W_2(\zeta)=W_1(\zeta) \int^\zeta  \frac{s^{-2}\rmd s}{W_1(s)^2}=
W_1(\zeta) \int^\zeta s^{-m-1}
 \frac{d s}{f(s)^2}.   
\end{equation}
Let us denote
\[
\frac{1}{f(\zeta)^2} = 1 + \sum_{k=1}^\infty g_{k}\zeta^k.
\]
Then, from~\eqref{eq:w2} it follows that the solution $W_2(\zeta)$ can be
written in the form
\begin{equation} 
\label{eq:w2f}
W_2(\zeta) = g_m W_1(\zeta) \ln\zeta  + \zeta^{\rho_-} V(\zeta),
\end{equation}
where $V(\zeta)$ is holomorphic in a neighbourhood of $\zeta=0$.  The
form of local monodromy depends on whether a logarithmic term is
present or not in the solution.  To check if it is present in our case,
we have to calculate if $g_3\neq 0$. It can be easily shown that
\[
g_3 = -2( 2 f_1^3  - 3 f_1 f_2 + f_3).
\]
The coefficients $f_i$, i=1,2,3 of the expansion \eqref{eq:w1} can be
computed directly (see e.g. \cite{Whittaker:65::}) and they are the
following
\begin{gather}
f_1 = \frac{1}{2\omega}, \qquad 
f_2= \frac{\omega^2(4k^2-1) +2(2-\xi)}{20\omega^2}, \\
f_3 = \frac{\omega^2(108k^2-47)+2(9-7\xi)}{360\omega^3}.
\end{gather}
One can check  that
\[
g_3 = \dfrac{(\omega^2-2\xi)}{9\omega^3}.
\]
Thus, if $\omega^2\neq 2\xi$ the logarithmic term in the solution
$W_2(\zeta)$ is present. Note that for $C<1$ the condition $\omega^2\neq
2\xi$ is equivalent to $2\xi\neq 3(1-C)$. Now,  let us consider a
small loop $\gamma$ encircling the singular point $\zeta=0$
counterclockwise.  The continuation of the matrix of the fundamental solutions
along this loop (under the assumption that $\omega^2\neq 2\xi$) gives rise
to the triangular monodromy matrix
\begin{equation}
\label{eq:monz}
\begin{bmatrix}
W_1(\zeta) & W_2(\zeta) \\
W_1'(\zeta) & W_2'(\zeta) 
\end{bmatrix} \xrightarrow[\gamma]{\qquad}
\begin{bmatrix}
W_1(\zeta) & W_2(\zeta) \\
W_1'(\zeta) & W_2'(\zeta) 
\end{bmatrix}
\begin{bmatrix}
1 & 2\pi i \\
0 & 1
\end{bmatrix}.
\end{equation}
This ends the proof. 
\end{proof} 
In the next lemma we show that for almost all values of the parameters
equation~\eqref{eq:normal} is not reducible, i.e., for it case 1 in
Lemma~\ref{lem:alg} does not occur. 
\begin{lemma}
\label{lem:redu}
  For $C\neq 1$ and $k\in(0,1)$ equation \eqref{eq:normal} is not
  reducible except for the case when
  \begin{equation}
   \xi= \frac{3}{2}(1-C), \qquad\text{and}\quad\omega^2=\dfrac{2}{2k^2-1}.
  \end{equation}
\end{lemma}
\begin{proof}
  To prove our Lemma we apply directly the first case of the Kovacic
  algorithm (see Appendix). First we consider the case $C<1$. All finite
  poles of $r(z)$ and infinity are of the second order.  Using
  the coefficients $a_i$, $i=1, \ldots, 4$ given by~\eqref{coeff} and
  the expansion~\eqref{nies} we obtain
\begin{equation}
\Delta_1=\Delta_2=
\Delta_3=\Delta_4=\dfrac12,\qquad \Delta_{\infty}=3,
\end{equation}
and thus
\begin{equation}
E_1=E_2=E_3=E_4=\left\{\dfrac14,\dfrac34\right\},\qquad E_\infty=\{-1,2\}.
\end{equation}
We proceed to the Second Step.  From the Cartesian product
$E=E_\infty\times\prod_{i=1}^4E_i$ we select these elements
$e=(e_\infty,e_1,e_2,e_3,e_4)\in E$ for which
\begin{equation}
d(e)=1-\left(e_\infty+\sum_{i=1}^4e_i\right)\in\mathbb{N}_0,
\end{equation}
where $\N_0$ denotes the set of non-negative integers.
In our case there exist seven elements of $E$ satisfying this condition
\begin{align*}
   e^{(1)} &= \left\{-1,\dfrac14,\dfrac14,\dfrac14,\dfrac14\right\}, 
   & d(e^{(1)})=1 ,&&\\
 e^{(2)} &= \left\{-1,\dfrac14,\dfrac14,\dfrac34,\dfrac34\right\}, 
   & d(e^{(2)})=0 ,&&\\
   e^{(3)} &= \left\{-1,\dfrac34,\dfrac34,\dfrac14,\dfrac14\right\}, 
   & d(e^{(3)})=0 ,&&\\
   e^{(4)} &= \left\{-1,\dfrac14,\dfrac34,\dfrac34,\dfrac14\right\}, 
   & d(e^{(4)})=0 ,&&\\
   e^{(5)} &= \left\{-1,\dfrac34,\dfrac14,\dfrac14,\dfrac34\right\}, 
   & d(e^{(5)})=0 ,&&\\
   e^{(6)} &= \left\{-1,\dfrac14,\dfrac34,\dfrac14,\dfrac34\right\}, 
   & d(e^{(6)})=0 ,&&\\
   e^{(7)} &= \left\{-1,\dfrac34,\dfrac14,\dfrac34,\dfrac14\right\}, 
   & d(e^{(7)})=0.
\end{align*}
   Now we pass to the Third Step of the Kovacic algorithm.  For each
   element $e\in E$ such that $d(e)\in\mathbb{N}_0$, we construct a
   rational function
\begin{equation}
  \label{eq:w}
    w=w(e)= \sum_{i=1}^4 \frac{e_i}{z-z_i}.
\end{equation}
Then we check if there exists a monic polynomial $P\in\mathbb{C}[z]$ of
degree $d(e)$ satisfying the equation
\begin{equation}
 \label{eq:Pol}
   P'' +2 w P' + (w' + w^2 - r) P = 0.
\end{equation}
If we find such polynomial, then equation~\eqref{eq:normal} has an
exponential solution $W=P\exp \int w$.

For $e^{(1)}$ we have $d(e^{(1)})=1$, thus we take $P=z+g$ and,
substituting $P$ to equation \eqref{eq:Pol}, we obtain the following
algebraic system
\begin{equation}
g\left[\omega^2(k^2-1)+\xi-1\right] = 0,\quad
  - 2\,g\,\omega - k^2\,\omega^2 + \xi -1 = 0,\quad
  \omega(g\omega+1) = 0.
\end{equation}
We note that  $\omega\neq 0$, and thus this system has one  solution
\[
g=-\dfrac{1}{\omega}, \qquad \xi=\dfrac{1}{2k^2-1}, \qquad
 \omega^2=\dfrac{2}{2k^2-1}.
\]
For $e^{(i)}$ $i=2,\ldots,7$ we have to find a monic polynomial of
degree zero satisfying \eqref{eq:Pol}, so we  put $P=1$.
 
For $e^{(2)}$ equation~\eqref{eq:Pol} yields
 \begin{equation}
 \omega^2+1-\xi=0,\qquad \omega=0,
 \end{equation}
but $\omega\neq 0$, so there is no solution of the above equations.

We have the same situation  for $e^{(3)}$ when \eqref{eq:Pol} gives
\begin{equation}
 \xi-1=0,\qquad\omega=0.
 \end{equation}
For $e^{(m)}$ with  $m=4,\dots, 7$ we obtain two equations of the form
\begin{equation}
  \omega^2(2k(k\mp ik')-3)+4(\xi-1)=0,\qquad
  \omega(\omega(k\mp ik')\pm 2)=0, 
\end{equation}
where the choice of signs depends on $m$.  The second equation cannot be
satisfied by a real $\omega\neq 0$ and $k\in(0,1)$.  This finishes the
proof for $C<1$. The proof for $C>1$ is similar.
\end{proof}
Combining the above two lemmas we have.
\begin{lemma}
\label{lem:main}
If $C\neq 1$ and $2\xi\neq {3}(1-C)$, then for $k\in(0,1)$ the
differential Galois group $\cG$ of~\eqref{eq:normal} is
$\mathrm{SL}(2,\C)$.
\end{lemma}
\begin{proof}
  In fact, under the given assumptions $\cG$ cannot be a triangular
  subgroup of $\mathrm{SL}(2,\C)$ by Lemma~\ref{lem:redu}. Under the
  same assumptions, by Lemma~\ref{lem:log}, we know that $\cG^0$
  contains a non-diagonalisable triangular matrix $M_\infty$. Thus case
  2 in Lemma~\ref{lem:alg} cannot occur as in this case $\cG^0$ is
  diagonal. By the same reason case 3 in Lemma~\ref{lem:alg} cannot
  occur as for a finite group $\cG$ the identity component $\cG^0$
  consists of the identity. Thus, we have
  $\cG=\cG^0=\mathrm{SL}(2,\C)$.
\end{proof}
As $ \mathrm{SL}(2,\C)$ is not Abelian, we have, as a direct
consequence of the above lemma, the following.
\begin{lemma}
\label{lem:mainn}
If $C\neq 1$ and $2\xi\neq {3}(1-C)$, then for $k\in(0,1)$
the complexified Hamiltonian system given by~\eqref{eq:Hp30} does not
admit an additional complex meromorphic first integral functionally
independent together with $H$ in a neighbourhood of phase curve
$\Gamma_k$.
\end{lemma}
However, as we mentioned the Hamiltonian system given by~\eqref{eq:Hp30}
is a subsystem of \eqref{eq:bel}, thus as a corollary we have the following 
theorem.
\begin{theorem}
\label{thm:mainc}
If $C\neq 1$, $A=B=1$, $\bL=(0,0,1)$ and $2\xi\neq {3}(1-C)$,
then for $k\in(0,1)$ the complexified system ~\eqref{eq:bel} considered on
$\cM^6$ does not admit an additional complex meromorphic first
integral functionally independent together with $H$ and $H_5$ in a
neighbourhood of the  phase curve $\Gamma_k$.
\end{theorem}
In the above Theorem the case $2\xi=3(1-C)$ is excluded. One can
suspect that for these values of the parameters our system is integrable.
Indeed, the lemma below shows that our suspicions are well justified
because a necessary condition for the integrability is satisfied.
\begin{lemma}
\label{lem:ab}
  If $2\xi= 3(1-C)$ then for all $k\in(0,1)$ the
  identity component of the differential Galois group
  of~\eqref{eq:normal} is Abelian.
\end{lemma}
\begin{proof}
We consider the case $C<1$. The proof for the  case $C>1$ is
similar. 

By Lemma~\ref{lem:redu} we know that for $2\xi= 3(1-C)$
equation~\eqref{eq:normal} is reducible only when
$\omega^2=2/(2k^2-1)$. As all exponents are rational and
equation~\eqref{eq:normal} is Fuchsian, in this case $\cG$ is a
proper subgroup of the triangular group so $\cG^0$ is Abelian.

For $\omega^2\neq 2/(2k^2-1)$ we show that case 2 of
Lemma~\ref{lem:alg} occurs.  To this end we apply the Kovacic
algorithm for this case.  Now sets $E_1,E_2,E_3,E_4$ and $E_\infty$
have the following forms
\begin{equation}
E_1=E_2=E_3=E_4=\{1,2,3\},\qquad E_\infty=\{-4,2,8\}.
\end{equation}
We have to find at least one monic polynomial $P\in\mathbb{C}[z]$ of
degree
\begin{equation}
d(e)=2-\dfrac12\left(e_\infty+\sum_{i=1}^4e_i\right),
\end{equation}
satisfying the differential equation
\begin{equation}
P'''+3wP''+(3w^2+3w'-4r)P'+(w''+3ww'+w^3-4rw-2r')P=0,
\label{eq3}
\end{equation}
where
 \begin{equation}
w=w(e)=\dfrac12 \sum_{i=1}^4 \frac{e_i}{z-z_i}.
\label{fer}
 \end{equation}
 We choose $e=(-4,1,1,1,1)$. Then $d(e)=2$, and we look for a
 polynomial of the second degree
 \begin{equation}
 P(z)=z^2+g_1z+g_2,
 \label{poly}
 \end{equation}
 satisfying \eqref{eq3}.  Substituting \eqref{poly} and \eqref{fer}
 into \eqref{eq3} we obtain the following system determining $g_1$ and
 $g_2$
 \begin{equation}
 \begin{split}
 &[(2k^2-3)\omega^2 +4(\xi-1)]g_1-4\omega g_2=0,\quad
 3\omega g_1+2\omega^2g_2+2(k^2+1-\xi)=0,\\
& \omega(\omega g_1+2)=0.
\end{split}
\end{equation}
If $\xi={\omega^2}/{2}$ then the above system has the following
solution
\begin{equation}
g_1=-\dfrac{2}{\omega},\qquad g_2=\dfrac{(1-2k^2)\omega^2+4}{2\omega^2}.
 \end{equation}
\end{proof}
%
%
%
%
\section{Real non-integrability}
\label{sec:rnon}
On $\cN$ system~\eqref{eq:bel} has four equilibria 
\[
s_\pm = (1,0,0,0,0,1,0,\mp 1,0), \qquad u_\pm =(1,0,0,0,0,1,0,0, \pm 1).
\]
These equilibria correspond to a fixed position of the satellite in
the orbital frame. For $s_\pm$ the symmetry axis is parallel to the
radius vector of the centre of mass of the satellite, and for $u_\pm$
the symmetry axis lies in the orbital frame and it is perpendicular to
the radius vector.

\begin{figure}[th]
\centering{ \includegraphics[scale=0.6]{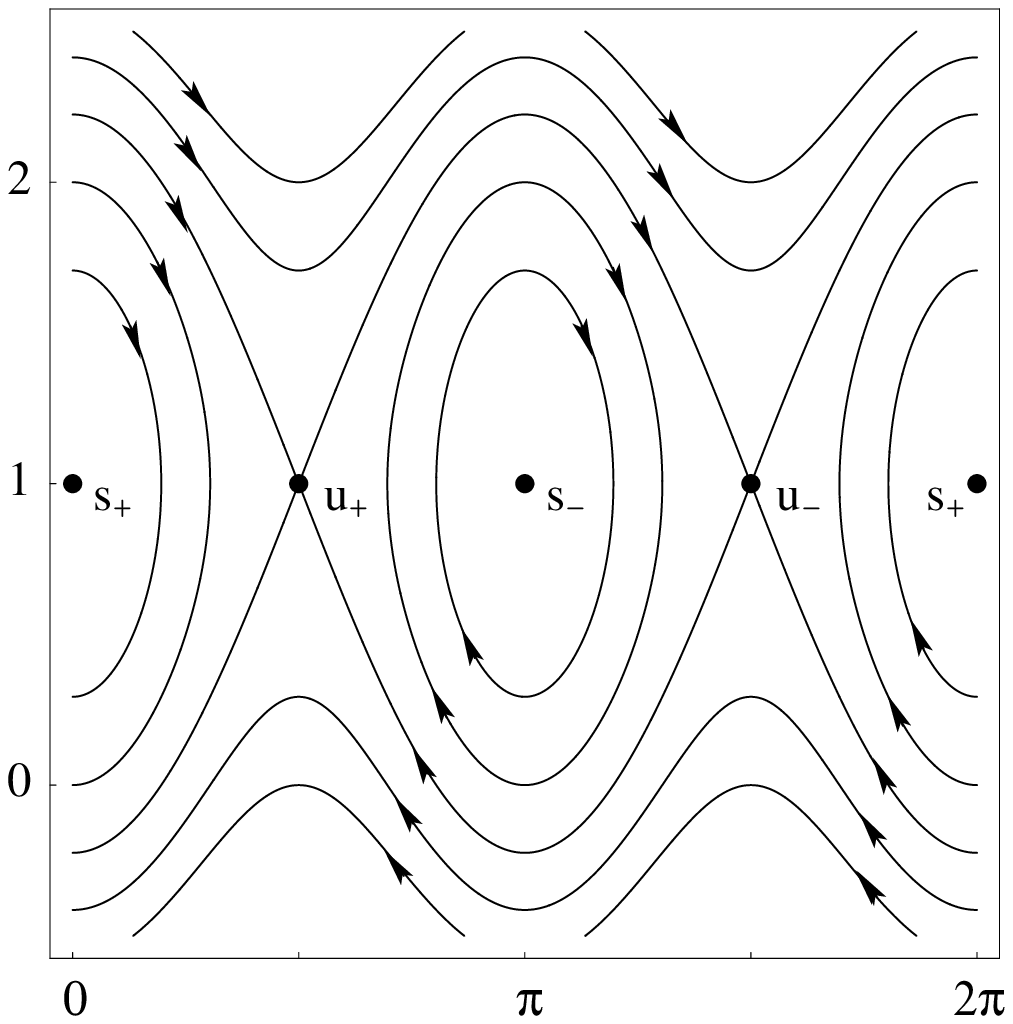}%
            \hspace{1cm} \includegraphics[scale=0.6]{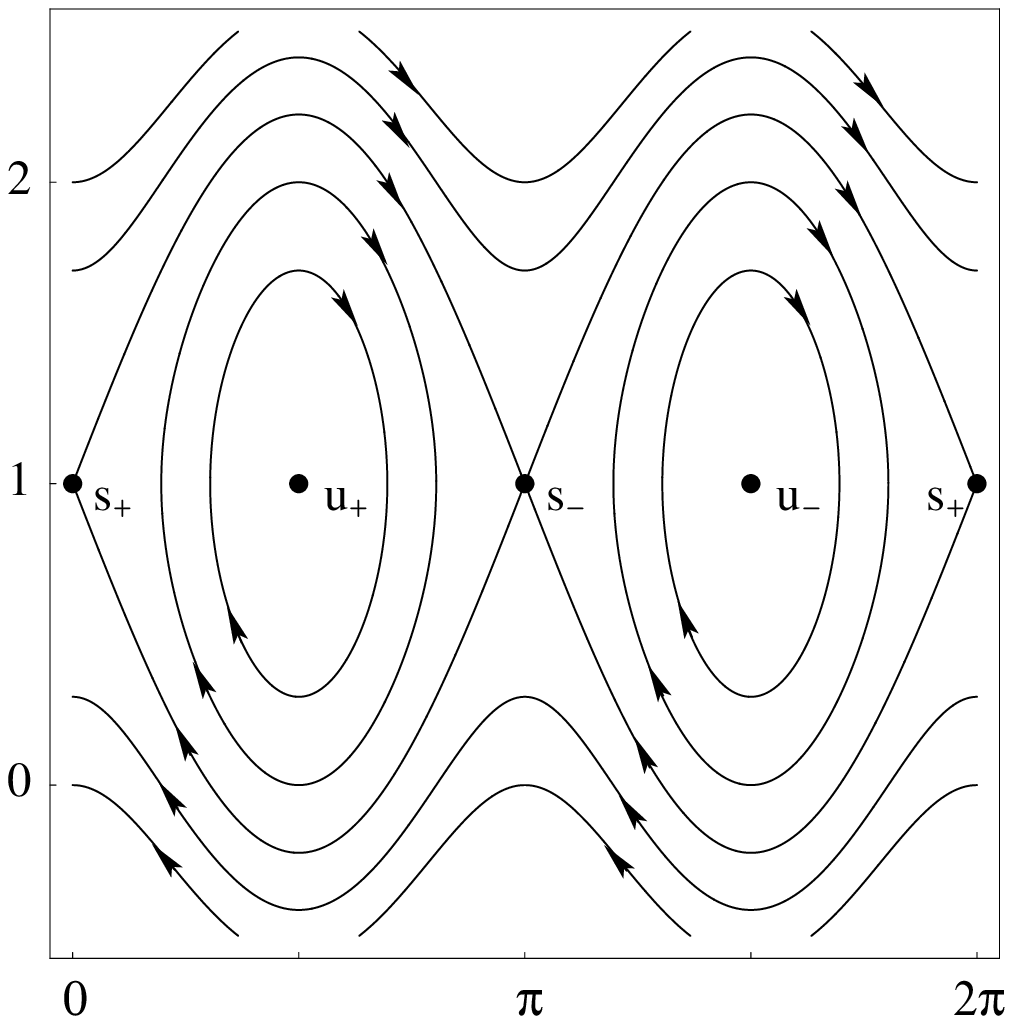} }
\caption{Phase portrait of planar oscillations in the $(\phi,M_1)$ plane. 
  The left panel corresponds to the case when $C>1$, and the right panel
  corresponds to the case when $C<1$.}
\label{fig:php}
\end{figure}

Let us restrict system~\eqref{eq:bel} to the invariant manifold
$\Pi=\cN\cap\cM^6$. Then, for the restricted system the equilibrium points
$u_\pm$ are hyperbolic if $C>1$; if $C<1$, then $s_\pm$ are hyperbolic.
See Figure~\ref{fig:php}.
%
%

We restrict further discussion to the case $C>1$. For real $t$ and
$0<k<1$ the solution $\Phi(t,k)$ defined by \eqref{eq:M1t}--
\eqref{eq:sBlC} corresponds to closed phase curves around the stable point
$s_+$. Closed real phase curves around $s_-$ are given parametrically by
$\Phi(t+T'(k)/2)$, $t\in\R$.  Let $\Gamma_1$ be the phase curve
corresponding to the solution given by \eqref{eq:M1t}--
\eqref{eq:sBlC} with $k=1$, i.e., $\Phi(t,1)$.  Then $\Gamma_1$
contains four components which are real phase curves corresponding to
real solutions heteroclinic to $u_\pm$.  Their union is $\Re \Gamma_1$,
and by $\Omega$ we denote the closure of $\Re \Gamma_1$.
\begin{lemma}
\label{lem:eps}
Let us assume that $C\neq 1$. Then for an arbitrary complex
neighbourhood $U\subset\Pi$ of $\Omega$ there exists $\epsilon>0$,
such that for $0<|k-1|<\epsilon$ the fundamental group
$\pi_1(\Gamma_k)$ of phase curve $\Gamma_k$ is generated by loops
lying in $U$.
\end{lemma}
\begin{proof}
  The periods $T(k)$ and $T'(k)$ of the solution $\Phi(t,k)$ are
  primitive and at the same time they are the minimal real and
  imaginary periods, respectively.  We choose the parallelogram of the
  fundamental periods as in Figure~\ref{fig:periods}. As a base point
  $x_0(k)\in\Gamma_k$ we choose $x_0(k)=\Phi_k(t_0(k))$ where
  $t_0(k)=T(k)/4$. Let us notice that from \eqref{eq:M1t}--
  \eqref{eq:sBlC} it follows that for $C>1$ we have
\begin{equation}
\label{eq:atok} 
M_1(t_0(k),k)= 1, \qquad S_2(t_0(k),k) = -k', \qquad  S_3(t_0(k),k)= k.  
\end{equation}
Now, we consider four loops 
\[
 \lambda(k), \;\lambda'(k),\; \gamma(k),\; \gamma'(k)  
:[0,1]\to\Gamma_k.
\]
The loops $ \lambda(k)$ and $\lambda'(k)$ correspond to the real and
imaginary periods, respectively (i.e. they correspond to the loops
$\alpha$ and $\alpha'$ in the  parallelogram of the periods, see
Figure~\ref{fig:periods} ).  
%
%
\begin{figure}[th]
\centering{ \includegraphics[scale=0.8]{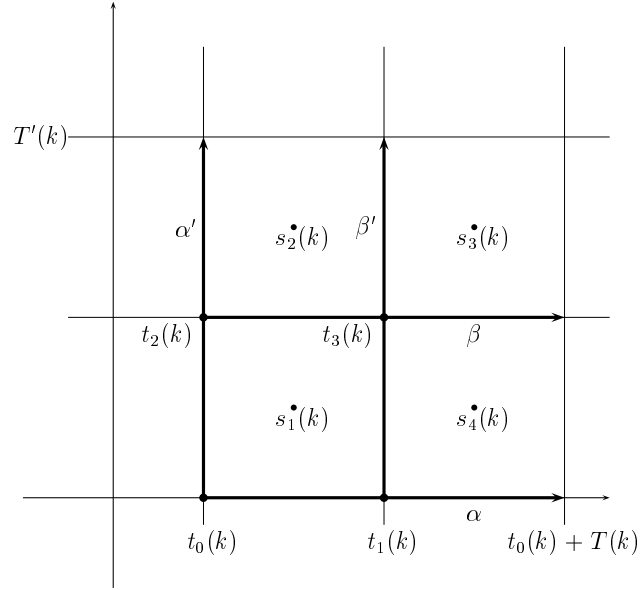} }
\caption{
  Parallelogram  of periods with chosen paths. The points marked by $t_i(k)$,
 with  $i=0,1,2,3$ are crossing points of the loops $\alpha$, $\alpha'$,
  $\beta$ and $\beta'$.}
\label{fig:periods}
\end{figure}

The loops $\gamma(k)$ and $\gamma'(k)$ corresponds to the `shifted' real
and imaginary periods, i.e. the  loops $\beta$ and $\beta'$ in the
 parallelogram of the periods.
\begin{remark}
\textsl{
  Above we use informal language. The correspondence between loops on
  $\Gamma(k)$ and paths on the complex time plane can be viewed as
  follows. The map
\[ 
   \C\ni t \longrightarrow \Phi(t,k)\in\Gamma(k),
\]
is a covering map. For a loop $\sigma(k)$ on $\Gamma(k)$ we obtain a
path $\hat\sigma(k)$ on $\C$ which is a lifting of $\sigma(k)$ with
respect to $\Phi(\cdot,k)$, i.e.  $\hat\sigma(k)$ is defined as such
curve for which
\[
 \sigma(k) = \Phi(\cdot,k)\circ \hat\sigma(k).
\]
}
\end{remark}
 These four loops cross at four common points
$x_l(k)=\Phi(t_l(k), k)$, $l=0,1,2,3$ where $t_1(k) = t_0(k)+T(k)/2$,
$t_2(k)=t_0(k) +T'(k)/2$ and $t_3(k)=t_0(k) +T(k)/2 + T'(k)/2$.
Moreover, we have
\begin{equation}
\label{eq:atik} 
\begin{split}
M_1(t_1(k))&= 1, \qquad S_2(t_1(k),k) = -k', \qquad  S_3(t_1(k),k)= -k, \\ 
M_1(t_2(k))&= 1, \qquad S_2(t_2(k),k) = +k', \qquad  S_3(t_2(k),k)= +k, \\ 
M_1(t_3(k))&= 1, \qquad S_2(t_3(k),k) = +k', \qquad  S_3(t_3(k),k)= -k. 
\end{split} 
\end{equation}
Thus, as $k$ tends to 1, the points $x_l(k)$ tend to $u_\pm$ and the loops
$\lambda(k)$ and $\gamma(k)$ approach $\Omega$.  We show that the loops
$\lambda'(k)$ and $\gamma'(k)$ tend to $u_\pm$. In fact, for $t=t_0(k)
+ \rmi \tau$, $\tau\in\R$ (i.e. along the loop $\alpha'$) from formulae
\eqref{eq:M1t}-- \eqref{eq:sBlC} we deduce that $S_2(t,k)$ and
$S_3(t,k)$ are real while $M_1(t,k)-1$ is purely imaginary. If we put
$M_1-1= \rmi \tilde M_1$ in \eqref{eq:quadrics} we obtain
\[
-\frac{1}{2}{\tilde M_1}^2 -\frac{1}{2} + 
\frac{3}{2}(S_2^2 + CS_3^2) = \frac{1}{2}\omega^2 k^2 +1, 
\qquad S_2^2 + S_3^2 = 1,
\]
and thus
\[
-{\tilde M_1}^2 +\omega^2 (1-S_2^2) = \omega^2k^2.
\]
It follows that for $k=1$
\[
{\tilde M_1}^2 +\omega^2 S_2^2 = 0, 
\]
but $\tilde M_1$ and $S_2$ are real, so we have $M_1-1=0$ and $S_2=0$,
i.e. the  loop $\lambda'(k)$ tends to $u_+$ as $k$ tends to 1. Similarly
we show that $\lambda'(k)$ tends to $u_-$ as $k$ tends to 1. 
%
%
\begin{figure}[th]
\centering{ \includegraphics[scale=0.8]{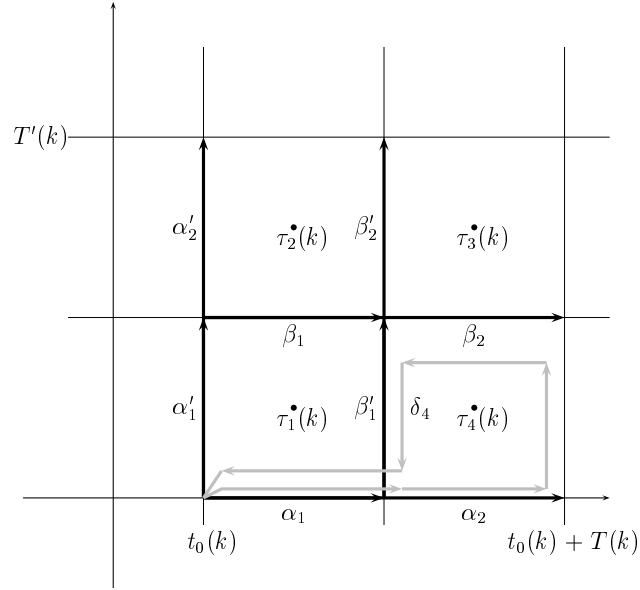} }
\caption{Loop $\delta_4$ is homotopic with loop $\alpha \cdot \alpha'_1
  \cdot \beta_2^{-1} \cdot \beta_1'^{-1} \cdot \alpha_1^{-1}$, where
  $\alpha = \alpha_1\cdot \alpha_2$.}
\label{fig:p2}
\end{figure}
Four points $x_l(k)$, $l=0,1,2,3$ divide four loops $ \lambda(k)$,
$\lambda'(k)$, $ \gamma(k)$ and $ \gamma'(k)$ into eight semi-loops $
\lambda_l(k)$, $\lambda_l'(k)$, $ \gamma_l(k)$ and $ \gamma_l'(k)$
which correspond to eight semi-loops $\alpha_l$, $\alpha_l'$,
$\beta_l$ and $\beta_l'$, $l=1,2$ in the  parallelogram of the periods.  Of
course we have $\lambda(k)=\lambda_1(k)\cdot\lambda_2(k)$, $\alpha
=\alpha_1\cdot\alpha_2$, etc.  We show that the fundamental group
$\pi(\Gamma(k), x_0(k))$ is generated by closed loops which are
appropriate compositions of these eight semi-loops.  The fundamental group
$\pi(\Gamma(k), x_0(k))$ is generated by homotopic classes of six
loops with the base point at $x_0(k)$: $\lambda(k)$, $\lambda'(k)$, and
four loops $\sigma_l(k)$ encircling the  singular points $s_l(k)$,
$l=1,2,3,4$. They satisfy the following condition:
\[
\sigma_1(k)\cdot\sigma_2(k)\cdot\sigma_3(k)\cdot\sigma_4(k) = 
\lambda(k)\cdot \lambda'(k)\cdot\lambda(k)^{-1}\cdot\lambda'(k)^{-1}.
\]
We show that the loop $\sigma_l(k)$ has the same homotopic class as an
appropriate composition of the semi-loops $\lambda_l(k)$, $\lambda_l'(k)$,
$ \gamma_l(k)$ and $ \gamma_l'(k)$. For example: 
\[
[\sigma_4(k)]= [\lambda(k)\cdot\lambda'_1(k)  \cdot\gamma_2(k)^{-1}  
\cdot\gamma_1'(k)^{-1}\cdot\lambda_1(k)^{-1}].
\]
Let $\delta_4$ be the loop encircling $\tau_4(k)$ and let
$\sigma_4(k)$ correspond to $\delta_4$. Then, we easily deduce that
$\delta_4$ has the same homotopic class as $\alpha \cdot \alpha'_1
\cdot \beta_2^{-1} \cdot \beta_1'^{-1} \cdot \alpha_1^{-1} $, 
see~Figure~\ref{fig:p2}.

Thus, we show that all generators of the fundamental group $\pi(\Gamma(k),
x_0(k))$ approach $\Omega$ as $k$ tends to 1.  
\end{proof} 
Now, we are ready to prove our main result. 
\begin{theorem}
\label{thm:maincc}
If $C\neq 1$, $A=B=1$, $\bL=(0,0,1)$,  and $2\xi\neq {3}(1-C)$,
then  system ~\eqref{eq:bel} considered on
$\cM^6$ does not admit an additional real  meromorphic first
integral functionally independent together with $H$ and $H_5$ in a
neighbourhood of the phase curve $\Gamma_1$.
\end{theorem}
\begin{proof}
  Let us assume that such meromorphic integral exists in a real
  neighbourhood of the phase curve $\Gamma_1$. Then we can extend it to a
  complex meromorphic first integral in a complex neighbourhood
  $\widetilde U$ of $\Gamma_1$. Then, by Lemma~\ref{lem:eps} we find
  $\Gamma_k$ with $k$ close to 1 such that its fundamental group is
  generated by loops which lie entirely in $\widetilde U$. From the
  Ziglin Lemma~\ref{lem:zig} it follows that the monodromy group of
  the NVE~\eqref{eq:nvet} possesses an invariant. But the  NVE~\eqref{eq:nvet}
  is a Fuchsian equation and thus if its monodromy group possesses an
  invariant, then its differential Galois group also possesses an
  invariant, see Theorem~3.17 in~\cite{Braider:96::}. However, by
  Lemma~\ref{lem:main} we show that the identity component of the
  differential Galois group of~\eqref{eq:normal}, and thus the
  identity component of the differential Galois group
  of~\eqref{eq:nvet}, is $\mathrm{SL}(2,\C)$. It follows that it does
  not possess an invariant, see Example~2.11(b)
  from~\cite{Braider:96::}. A contradiction finishes the proof.
\end{proof}
%
%
\section{Comments and Remarks}
Although, as it is commonly believed, most systems are not integrable
and integrable systems are extremely rare, the example considered in
this paper shows that to prove the non-integrability one has to use
rather involved techniques. Nevertheless, a proof of
non-integra\-bility of a system gives, in some sense, a negative
result---the true aim is to find a non-trivial integrable system.
From this point of view, the reader can wonder why we did not
investigate carefully the case of the  parameter values $2\xi = 3(1-C)$
for which the necessary conditions for integrability are satisfied. As
a matter of fact, for some time we believed that for these parameter
values the system is integrable. With the help of the computer algebra
we tried to find a polynomial or rational first integral of the system
 but we failed. For  fixed values of $C$ we numerically generated the
Poincar\'e cross sections of the system which evidently showed that the
system is not integrable. Thus, our conjecture is that the system also
is non-integrable for the case $2\xi = 3(1-C)$. An analytic proof of this
fact needs a separate investigation.

For $\xi=0$ Theorem~\ref{thm:maincc} tells us that the problem of a
symmetric rigid satellite in a circular orbit is not integrable for
all values of $C\in(0,2)$ except $C=1$. This problem was also
investigated in \cite{Maciejewski:01::i,Maciejewski:01::j,Audin:02::}
where a proof of the same fact is given. However, in all these
references as a particular solution a heteroclinic orbit was chosen and
instead of transformation~\eqref{eq:ttoz} another one was used. 
This leads to a more complicated form  of the  NVE.  
\section{Acknowledgements}
We would like to thank Delphine Boucher, Juan J.~Morales-Ruiz,
Jacques-Arthur Weil, Carles Sim\'o, Michael F.~Singer and Felix Ulmer for
discussions and help which allowed  us to understand many topics related
to this work. As usual, we thank Zbroja not only for her linguistic help.
For the second author this research has been supported by a Maria
Curie Fellowship of the European Community programme Human Potential
under contract number HPMF-CT-2002-02031.

\section*{Appendix}
Let us consider a linear second order differential equation with rational 
coefficients
\begin{equation}
\label{eq:agl}
w'' + p(z) w' + q(z) w =0, \qquad p(z), q(z)\in \C(z).
\end{equation}  
A point $z=c\in\C$ is a singular point of this equation if it is a
pole of $p(z)$ or $q(z)$. A singular point is a \emph{regular
  singular} point if at this point $\tilde p(z)=(z-c)p(z)$ and
$\tilde q(z)=(z-c)^2q(z)$ are holomorphic. An \emph{exponent} of
equation~\eqref{eq:agl} at point $z=z_0$  is a solution of the
\emph{indicial equation}
\[
   \rho(\rho-1) +p_0\rho + q_0 = 0, \qquad p_0=\tilde p(c), \quad 
  q_0=\tilde q(c). 
\]
 After changing the dependent variable $z\to
  1/z$ equation~\eqref{eq:agl} reads
\begin{equation}
\label{eq:infagl}
w'' + P(z) w' + Q(z) w =0, \qquad 
P(z)=\frac{1}{z^3}p\left(\frac{1}{z}\right) +\frac{2}{z}, \qquad 
Q(z)= \frac{1}{z^4}q\left(\frac{1}{z}\right) .
\end{equation}  
We say that the point $z=\infty$ is a singular point for
equation~\eqref{eq:agl} if $z=0$ is a singular point
of~equation~\eqref{eq:agl}. Equation~\eqref{eq:agl} is called
\emph{Fuchsian} if all its singular points (including infinity) are
regular, see \cite{Whittaker:65::,Ince:44::} 

If one (non-zero) solution $w_1$ of  equation~\eqref{eq:agl} is
Liouvillian, then all its solutions are Liouvillian.  In fact, the
second solution $w_2$, linearly independent from  $w_1$, is given by
\[
w_2 = w_1\int \frac{1}{w_1^2}\exp\left[-\int p\right].
\]
Putting 
\begin{equation}
\label{eq:wtoy}
   w = y \exp\left[\frac{1}{2}\int p \right]
\end{equation}
into equation~\eqref{eq:agl} we obtain its \emph{reduced form}
\begin{equation}
\label{eq:gso}
 y''=r(z) y, \qquad r(z) = -q(z) + \frac{1}{2}p'(z)  + \frac{1}{4}p(z)^2.
\end{equation} 
This change of variable does not affect the Liouvillian nature of the
solutions.  For equation~\eqref{eq:gso} its differential Galois group
$\cG$ is an algebraic subgroup of $\mathrm{SL}(2,\C)$. The following
lemma describes all possible types of $\cG$ and relates these types to
forms of solution of \eqref{eq:gso}, see
\cite{Kovacic:86::,Morales:99::}.
\begin{lemma}
\label{lem:alg}
Let $\cG$ be the differential Galois group of equation~\eqref{eq:gso}.
Then one of four cases can occur.
\begin{enumerate}
\item $\cG$ is conjugated with a subgroup of the triangular group
\[
\cT = \left\{ \begin{bmatrix} a & b\\
                                0 & a^{-1}
                      \end{bmatrix}  \; \biggl| \; a\in\C^*, b\in\right\} ;
\]
 in this case equation
  \eqref{eq:gso}  has an exponential solution,
\item $\cG$ is conjugated with a subgroup of 
\[
D^\dag= \left\{ \begin{bmatrix} c & 0\\
                                0 & c^{-1}
                      \end{bmatrix}  \; \biggl| \; c\in\C^*\right\} \cup 
                      \left\{ \begin{bmatrix} 0 & c\\
                                c^{-1} & 0
                      \end{bmatrix}  \; \biggl| \; c\in\C^*\right\}, 
\]
  in this case equation
  \eqref{eq:gso} has a  solution of the form $y=\exp\int \omega$, where
  $\omega$ is algebraic over $\C(z)$ of degree 2,
\item $\cG$ is primitive and finite; in this case all
  solutions of equation~\eqref{eq:gso} are algebraic, 
  
\item $\cG= \mathrm{SL}(2,\C)$ and equation \eqref{eq:gso}
  has no Liouvillian solution.
\end{enumerate}
\end{lemma}
When the first case occurs we say that equation~\eqref{eq:gso} is
\emph{reducible}.  

The Kovacic algorithm~\cite{Kovacic:86::} allows to decide if an
equation of the form~\eqref{eq:gso} possesses a Liouvillian solution.
Applying it we also obtain information about the differential Galois
group of this equation. Now, beside the original formulation of this
algorithm~\footnote{On the web page
  \texttt{http://members.bellatlantic.net/~jkovacic/lectures.html} the
  reader will find lecture notes of J.J. Kovacic which contain an
  extended description of the algorithms with many remarks and
  comments concerning recent works on the subject.}  we have its
several versions and improvements and extensions to higher order
equations~\cite{Duval:89::,Duval:92::,Ulmer:96::,Singer:93::,Singer:95::,Hoeij:98::,Boucher:00::}

Here we present a part of the Kovacic algorithm which allows to decide
whether ~\eqref{eq:gso} possesses a solution of the form $y=\exp\int
\omega$, where $\omega$ is algebraic over $\C(z)$ of degree 1 or 2,
or, in other words it gives an answer whether
for equation~\eqref{eq:gso} case 1 or 2 in Lemma~\ref{lem:alg} can
occur. We used this part of the algorithm in Lemma~\ref{lem:redu} and
Lemma~\ref{lem:ab}. As our NVE is Fuchsian, we present the algorithm
adopted for a Fuchsian equation because it is simpler than for a
general case.

We write $r(z)\in\C(z)$ in the form
\begin{equation*}
\label{eq:rst}
r(z) = \frac{s(z)}{t(z)}, \qquad s(z),\, t(z) \in \C[z],
\end{equation*}
where $s(z)$ and $t(z)$ are relatively prime polynomials and $t(z)$ is
monic.  The roots of $t(z)$ are poles of $r(z)$. We denote  $\Sigma':=
\{ c\in\C\,\vert\, t(c) =0 \}$ and  $\Sigma:=\Sigma'\cup\{\infty\}$.  The
order $\ord(c)$ of $c\in\Sigma'$ is equal to the multiplicity of $c$
as a root of $t(z)$, the order of infinity is defined by
\[ 
\ord(\infty):= \max(0, 4+\deg s - \deg t).
\]
As we assumed,  equation~\eqref{eq:gso} is Fuchsian, so we have
$\ord(c)\leq 2$ of $c\in\Sigma$. For each $c\in\Sigma'$ we have the
following expansion
\[
r(z) = \frac{a_c}{(z-c)^2} + O\left( \frac{1}{z-c}\right),
\]
and we define $\Delta_c = \sqrt{1+4a_c}$. For infinity we have

\begin{equation*}
r(z)=\dfrac{a_\infty}{z^2}+O\left(\dfrac{1}{z^3}\right),
\end{equation*}
and we define $\Delta_\infty = \sqrt{1+4a_\infty}$. 

Now we describe the Kovacic algorithm for the two cases
mentioned.\\[\bigskipamount]
\noindent
\textsc{Case I}

\noindent
\textbf{Step I.}
For each $c\in\Sigma'$ such that $\ord{c}=1$ we define  $E_c=\{1\}$;
if  $\ord{c}=2 $  
\[
  E_c := \left\{ \frac{1}{2}\left(1+\Delta_c\right),
   \frac{1}{2}\left(1-\Delta_c\right)\right\}. 
\] 
If $\ord{(\infty)}<2$ we put $E_\infty=\{ 0,1\}$; if
$\ord{(\infty)}=2$ we define
\[
  E_\infty := \left\{ \frac{1}{2}\left(1+\Delta_\infty\right), 
  \frac{1}{2}\left(1-\Delta_\infty\right)\right\}. 
\] 
\textbf{Step II.} For each element $e$ in the Cartesian product
\[
E := E_\infty\times \prod_{c\in\Sigma'}E_c,
\]
we compute 
\[
  d(e) := 1 - \sum_{c\in\Sigma}e_c. 
\]
We select those elements $e\in E$ for which $d(e)$ is a non-negative
integer.  If there are no such elements equation~\eqref{eq:gso} does
not have an exponential solution and the algorithm stops here. 

\noindent 
\textbf{Step III.} For each element $e\in E$ such that
$d(e)=n\in\N_0$ we define\[ \omega(z) =
\sum_{c\in\Sigma'}\frac{e_c}{z-c},
\]
and we search for a monic polynomial $P=P(z)$ of degree $n$ satisfying the
following equation
\[ 
P'' + 2\omega(z)P' +( \omega'(z)+ \omega(z)^2 -r(z))P =0.
\] 
If such  polynomial exists, then equation~\eqref{eq:gso} possesses an
exponential solution of the form $y=P\exp\int\omega$, if not,
equation~\eqref{eq:gso} does not have an exponential solution. \\[\bigskipamount]
\noindent
\textsc{Case II}

\noindent
\textbf{Step I.}
For $c\in\Sigma'$ such that $\ord{c}=1$ we define  $E_c=\{4\}$;
if  $\ord{c}=2 $  
\[
  E_c := \left\{ 2 , 2(1+\Delta_c),
   2(1-\Delta_c)\right\}\cap\Z. 
\] 
If $\ord{(\infty)}<2$ we put $E_\infty=\{ 0,2,4\}$; if
$\ord{(\infty)}=2$ we define
\[
  E_\infty := \left\{2,  2(1+\Delta_\infty), 
  2(1-\Delta_\infty)\right\}\cap\Z. 
\] 
\textbf{Step II.} If the Cartesian product 
\[
E := E_\infty\times \prod_{c\in\Sigma'}E_c,
\]
is empty then case 2 cannot occur and algorithm stops here. If it is
not, then for $e\in E$ we compute
\[
  d(e) := 2 - \frac{1}{2} \sum_{c\in\Sigma}e_c. 
\]
We select those elements $e\in E$ for which $d(e)$ is a non-negative
integer.  If there are no such elements  case 2 cannot occur and
algorithm stops here.

\noindent 
\textbf{Step III.} For each element $e\in E$ such that
$d(e)=n\in\N_0$ we define
\[ 
\theta=\theta(z) = \frac{1}{2}
\sum_{c\in\Sigma'}\frac{e_c}{z-c},
\]
and we search for a monic polynomial $P=P(z)$ of degree $n$ satisfying the
following equation
\[ 
P''' + 3\theta P'' +(3 \theta^2 + 3\theta' -4r)P' + 
(\theta'' + 3 \theta\theta' + \theta^3 -4 r\theta - 2r')P =0.
\] 
If such a polynomial exists then equation~\eqref{eq:gso} possesses a
 solution of the form $y=\exp\int\omega$, where 
\[
\omega^2 - \psi\omega +\frac{1}{2}\psi' + \frac{1}{2}\psi^2 - r =0, \qquad 
\psi = \theta + \frac{P'}{P}.
\]
If we do not find such  polynomial, then case 2 in Lemma~\ref{lem:alg}
cannot occur.  
\bibliographystyle{plain}
\bibliography{mathreva,books,ham,bogoya,ajm,ziglin,dgt,oldies,tsygvintsev.bib}
\end{document}